\documentclass[preprint,pre,oneside,floatfix]{revtex4}

\usepackage{amsmath}
\usepackage{amssymb}
\usepackage{graphicx}
\usepackage{dcolumn}
\usepackage{bm}
\usepackage{color}

\def\dps{\displaystyle}

\def\Eq#1{(\ref{eq:#1})}

\def\epsilon{\varepsilon}
\def\theta{\vartheta}
\def\rho{\varrho}

\def\vec#1{\mathbf{#1}}

\begin{document}

%===============================================================================

\title{Electrostatic interaction between colloidal particles trapped at an electrolyte interface}

\author{Arghya Majee}
\email{majee@is.mpg.de}
\author{Markus Bier}
\email{bier@is.mpg.de}
\author{S. Dietrich}
\affiliation
{
   Max-Planck-Institut f\"ur Intelligente Systeme, 
   Heisenbergstr.\ 3,
   70569 Stuttgart,
   Germany, 
   and
   Institut f\"ur Theoretische Physik IV,
   Universit\"at Stuttgart,
   Pfaffenwaldring 57,
   70569 Stuttgart,
   Germany
}

\date{\today}

\begin{abstract}
The electrostatic interaction between colloidal particles trapped at the interface between two immiscible electrolyte
solutions is studied in the limit of small inter-particle distances. Within an appropriate model exact analytic expressions 
for the electrostatic potential as well as for the surface and line interaction energies are obtained. They demonstrate that 
the widely used superposition approximation, which is commonly applied to large distances between the colloidal particles, 
fails qualitatively at small distances and is quantitatively unreliable even at large distances. Our results contribute to an 
improved description of the interaction between colloidal particles trapped at fluid interfaces.  
\end{abstract}

\maketitle

%===============================================================================

\section{Introduction}

Colloidal particles, trapped at fluid interfaces by adsorption energies much larger than the
thermal energy, can form effectively two-dimensional colloidal monolayers \cite{Pie80}. During the last two decades these systems 
have received significant attention both in basic research as well as in applied sciences.
On one hand, these monolayers serve as model systems for studying effective interactions, phase behaviors, structures, and the 
dynamics of condensed matter in reduced dimensionality \cite{Joa01,Din02,Lou05,Che11,Wan12,Ers13,Mao13}.
On the other hand, self-assembled colloidal monolayers find applications in optical devices, molecular electronics,
emulsion stabilization processes, and as templates in the fabrication of new micro- and nanostructured materials. 
Therefore, a reliable description of the lateral inter-particle interaction at all distances $r$, which governs the structure formation of colloids
at fluid interfaces, is of primary importance.

In his pioneering work Pieranski \cite{Pie80} showed that the electrostatic {\it repulsion} of charged colloids at such interfaces is 
dominated by a long-ranged dipole-dipole interaction, due to an asymmetric counterion distribution in the two adjacent media, in addition to
the screened Coulomb interaction also present in bulk systems. 
Later both the power-law and the exponential contributions have been calculated within the framework of linearized Poisson-Boltzmann 
theory assuming point-like particles \cite{Hur85}. 
It turned out that, whereas the interaction energy for charged particles always decays asymptotically $\propto 1/r^3$, the prefactor depends on whether
the interaction originates from charges on the polar \cite{Pie80,Par08} or on the apolar \cite{Ave00,Ave02} side of the fluid 
interface. 
In addition there are experimental indications of an {\it attractive} long-ranged lateral interaction which cannot be interpreted
in terms of a van der Waals force \cite{Sta00,Nik02}.
Attempts were made to explain it in terms of a deformation-induced capillary interaction, but a complete and final picture has 
not yet been reached \cite{For04,Oet051,Oet052,Wue05}.
Here, we focus on the electrostatic contribution to the interaction. 

Whereas Pieranski's work has been extended in numerous directions, almost all subsequent studies have discussed
exclusively the case of colloidal particles being far away from each other. 
In this asymptotic limit the superposition approximation has been assumed to be reliable, according to which one approximates the actual 
electrostatic potential (or interfacial deformation) for a pair of particles by the sum of the potentials 
(or deformations) of the two single particles.
However, for a dense system or during aggregation, particles can come close to each other such that this superposition approximation
is no longer justified.
For the deformation induced attractive part of the interaction, the validity of this approximation has been 
discussed for both large \cite{Oet051,Wue05,Dom07} and small \cite{He13} separations.
But so far for the repulsive electrostatic interaction no investigations of small-distance deviations from the superposition 
approximation have been reported, although a systematic multipole expansion of the electrostatic potential around a 
single inhomogeneously charged particle trapped at an interface is available \cite{Dom08}.

\begin{figure}[!t]
   \includegraphics[width=8cm]{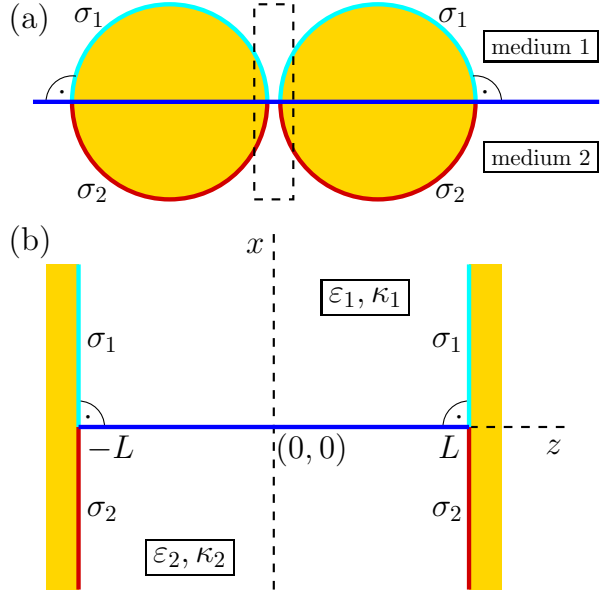}
   \caption{(a) Cross section of two identical spherical particles trapped at a fluid interface (horizontal blue line) close to each other and with
        contact angle $90^\circ$. (b) Magnified view of the boxed region in (a).
        The two adjacent fluids (``1'', located at $x>0$, and ``2'', located at $x<0$) forming the interface have permittivities $\varepsilon_1$, 
        $\varepsilon_2$ and inverse Debye lengths $\kappa_1$, $\kappa_2$, respectively. 
        Since the surface-to-surface distance between the particles is small compared to their radii, the particle surfaces can be approximated by 
        planes located at $z=\pm L$ which carry charge densities $\sigma_1$ and $\sigma_2$ at the surfaces in contact with fluid ``1'' and ``2'', 
        respectively. According to the model the fluid structures vary steplike at the surfaces and at the interface.
}
   \label{fig:1}
\end{figure}

\begin{figure}[!t]
   \includegraphics[width=7cm]{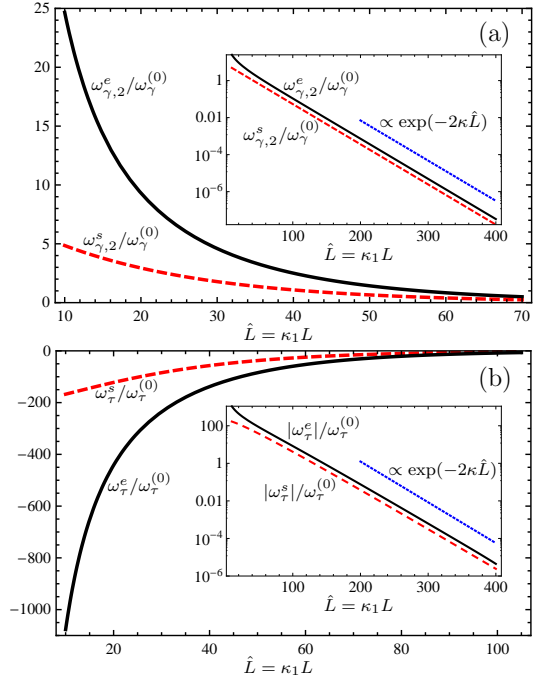}
   \caption{(a) Comparison between the exact expression (superscript ``e'', black solid lines, see Eq.~\Eq{m4}) and the corresponding 
        superposition approximation (superscript ``s'', red dashed lines, see Eq.~\Eq{m5}) of the surface interaction energy $\omega_{\gamma,2}(L)$ per 
        total surface area of contact between the walls and medium ``2'' in units of 
        $\omega^{(0)}_\gamma=\sigma_1^2/(\kappa_1\varepsilon_1)$ as a function of $\hat{L}=\kappa_1L$.
        Typical experimental values for the parameter ratios $\kappa=\kappa_2/\kappa_1=0.025$, $\varepsilon=\varepsilon_2/\varepsilon_1=0.025$, and 
        $\sigma=\sigma_2/\sigma_1=0.1$ have been chosen for the plots \cite{Sta00, Nik02, Dan04, Che09, Kes93}. Obviously $\omega^e_{\gamma,2}(L)$ and 
        $\omega^s_{\gamma,2}(L)$ 
        differ significantly at small distances, but even in the limit of large wall separations the superposition approximation is 
        too small by a factor of $2$ (see the offset between the two curve in the inset). A similar deviation is
        obtained for $\omega_{\gamma,1}(L)$, but due to its very small magnitude ($\approx10^{-10}\times\omega_{\gamma,2}(L)$, for the 
        above parameter choices) it is not shown here (see Fig.~\ref{fig:4si} in the appendix).
        (b) Comparison of the exact expression (superscript ``e'', black solid lines) and the superposition 
        approximation (superscript ``s'', red dashed lines) of the effective line interaction energy $\omega_\tau(L)$ per total length of the
        three-phase contact lines between media ``1'' and ``2'' and the walls in units of $\omega_\tau^{(0)}=\sigma_1^2/(\kappa_1^2\varepsilon_1)$
        as a function of $\hat{L}$ (see appendix~\ref{app:C} for explicit expressions).
        In addition to the same parameters $\sigma$, $\epsilon$, and $\kappa$ as in panel (a) the Donnan potential (Galvani potential 
        difference) $\Phi_D/\Phi_1^{(0)}=1.3$ is used. As for the surface interaction potential in panel (a), the superposition approximation of the 
        line interaction potential deviates
        qualitatively from the exact result at small wall separations and its absolute value at large distances is too small by a factor of $2$.}
   \label{fig:2}
\end{figure}

Here, we assess the quality of the superposition approximation for the electrostatic interaction between two colloidal 
particles floating close to each other at an electrolyte interface by considering a simplified problem (see Fig.~\ref{fig:1}) which offers
the possibility to obtain exact analytic expressions. Accordingly,
first, the interface is assumed to be planar, i.e., no deformations of the fluid interface are considered, which are
typically of the order of nanometers for micron-sized particles \cite{Sta00,Nik02,Law13}.
Second, due to the small particle-particle distances to be studied, the curvature of the colloidal particles is ignored in the
spirit of a Derjaguin approximation \cite{Rus89} by considering the effective interaction between two charged, 
planar, and parallel walls. 
Third, a liquid-particle contact angle of $90^\circ$ is assumed; this value is encountered for actual systems \cite{Mas10}.
We have derived an exact analytic expression for the electrostatic potential of this model within linearized Poisson-Boltzmann theory,
which is then used to calculate the surface interaction energies per total surface area and the line interaction energy per 
total length of the two three-phase contact lines (Fig.~\ref{fig:1}). 
The main result is the observation of significant deviations between the exact values of these quantities and those obtained within 
the superposition approximation, both at small and even at large distances (see Fig.~\ref{fig:2}). 

%===============================================================================

\section{Electrostatic potential} Consider a three-dimensional Cartesian coordinate system such that the two charged planar 
walls, which mimic the colloidal particles, are located at $z=\pm L$ and the fluid interface is at $x=0$ (Fig.~\ref{fig:1}(b)).
The electrolyte solution present at $x>0$ ($x<0$) is denoted as medium ``1'' (``2''). For simplicity here we consider binary monovalent electrolytes only, i.e., there
are only two ionic species of opposite sign like $\text{Na}^+$ and $\text{Cl}^-$.
Generically the ions and the molecules are coupled such that the molecular and ion number densities vary on the scale of the bulk 
correlation length which is much smaller than the Debye length which sets the length scale for the variation of the charge density \cite{Bie12}.
Thus the number densities in both media vary only close to the walls or to the fluid interface at distances of 
the order of the bulk correlation length, which, away from critical points, is of the order of the size of the fluid molecules and 
of the ions and falls below the length scale to be considered here.
Accordingly, the permittivity $\varepsilon_1$ ($\varepsilon_2$) and the inverse Debye length $\kappa_1$ ($\kappa_2$) in medium
``1'' (``2'') are uniform where $\kappa_i=(2I_ie^2/(\varepsilon_ik_BT))^{1/2}$, $i\in\{1,2\},$ with bulk ionic strength $I_i$ (which is
the bulk number density of each ionic species in medium $i$), Boltzmann constant $k_B$, temperature $T$, and elementary charge $e>0$.
The two walls are assumed to be chemically identical such that the surface charge densities at both half-planes in contact with medium
``1'' (``2'') are given by $\sigma_1$ ($\sigma_2$).
The local charge density of the ions is \emph{not} uniform in media ``1'' or ``2'' because this quantity varies on the
scale of the Debye lengths, which are typically much larger than molecular sizes.
Since the slab formed by the two walls at $z=\pm L$ is a model of the space in between two colloidal particles trapped at the
fluid interface, it is appropriate to describe the ions within a grand canonical ensemble, the reservoirs of which are given by the
bulk electrolyte solutions far away from the fluid interface.
Within a simple density functional theory, which (i) considers uniform solvents in the upper and the lower half space, (ii) assumes 
low ionic strength in the bulk (which 
facilitates the description of the ions as point-like particles), and (iii) describes deviations of the ion densities from the bulk ionic 
strengths only up to quadratic order, one derives the linearized Poisson-Boltzmann (PB) equation $(\Delta-\kappa_i^2)\Phi_i=0$ to be 
fulfilled by the electrostatic potential $\Phi_i(x,z)$ in medium $i\in\{1,2\}$.
The corresponding boundary conditions are: (i) the electrostatic potential should remain finite for 
$x\rightarrow\pm\infty$, (ii) the electrostatic potential and the normal component of the electric displacement field at the 
fluid interface should be continuous, i.e., $\Phi_1(x=0^+,z)=\Phi_2(x=0^-,z)$ and $\varepsilon_1\partial_x\Phi_1(x=0^+,z) = 
\varepsilon_2\partial_x\Phi_2(x=0^-,z)$, and (iii) due to global charge neutrality the normal component of electric displacement 
field at the walls correspond to the surface charge densities, i.e., $\varepsilon_i\partial_z\Phi_i(x,z=\pm L)=\pm\sigma_i$.
It is important to note that in our model the fluids are confined to the space between the two walls such that outside the fluid slab
the electric field vanishes. 

In order to determine the electrostatic potential we first split the whole problem into three sub-problems (see appendix~\ref{app:A}): (i) 
only the fluid interface is present in the absence of any walls, (ii) two charged 
walls with homogeneous surface charge densities $\sigma_1$
and the uniform medium ``1'' in between, and (iii) two charged walls with homogeneous surface charge 
densities $\sigma_2$ and the uniform medium ``2'' in between. 
By adding the solution of problem (ii) and the solution of problem (i) for the upper half-space 
and by adding the solution of problem (iii) and the solution of problem (i) for the lower half-space,
one obtains potentials in the two media which satisfy all the boundary conditions listed above except the continuity of the 
potential at the interface. In order to fulfill also the latter one, we construct a correction function which (i) is a solution of 
the linearized PB equation, (ii) keeps all boundary conditions unchanged which are already satisfied, and (iii) leads to 
continuity of the potential at the interface. This can be achieved by means of 2D Fourier transform or Fourier series expansions \cite{Sti61}. 
The final expression for the {\it{e}}xact electrostatic potential (denoted by superscript ``e'') reads
\begin{align}
   \Phi_i^e&(x,z)=\!\Phi_{bi}\!+\sum\limits_{j\in\{1,2\}}^{j\neq i}
                  \frac{(-1)^j\kappa_j\varepsilon_j\Phi_D}{\kappa_1\varepsilon_1+\kappa_2\varepsilon_2}
                  e^{-\kappa_{i}\lvert x\rvert}\notag\\
                 &+\Phi_i^{(0)}\frac{\cosh(\kappa_iz)}{\sinh(\kappa_iL)}\!
                  +\!\!\sum\limits_{j\in\{1,2\}}^{j\neq i}\frac{C_{ij}^{(0)}(L)e^{-a_i^{(0)}(L)\lvert x\rvert}}{2}\notag\\
                 &+\sum\limits_{j\in\{1,2\}}^{j\neq i}\sum\limits_{n=1}^\infty C_{ij}^{(n)}(L)e^{-a_i^{(n)}(L)\lvert x\rvert}\cos\left(\frac{n\pi z}{L}\right),
   \label{eq:m1}
\end{align}
where the explicit dependences of $\Phi_i^{(0)}$, $a_i^{(n)}(L)$, and $C_{ij}^{(n)}(L)$ on $n$, $L$, and the type of media $i$ and $j$ are given 
in appendix~\ref{app:A}. 
The electrostatic bulk potential $\Phi_{bi}$ is defined as $\Phi_{b1}=0$ and $\Phi_{b2}=\Phi_D$, with the Donnan potential 
(or Galvani potential difference \cite{Bag06}) $\Phi_D$ between medium ``2'' and medium ``1'', which originates from the differences 
of the solubilities of the ions in the two media \cite{Bie08}.

The first two terms on the right-hand side of Eq.~(\ref{eq:m1}) together represent the effect of the fluid interface in the absence
of walls (sub-problem (i)) which corresponds to the limit $L\rightarrow\infty$ at any fixed position $z$. 
The third term describes the electrostatic potential of two uniformly and equally 
charged walls in the presence of a uniform electrolyte solution in between (sub-problem (ii) or (iii)). According to Eq.~\Eq{m1},
up to the constant $\Phi_{bi}$, $\Phi_i^e(x,z)$ reduces to the third term in the limit $\lvert x\rvert\rightarrow\infty$, i.e.,
far away from the fluid interface.
The fourth and the fifth term in Eq.~(\ref{eq:m1}) correspond to the correction function which describes the contact of the walls
with the fluid interface. 
Due to the symmetry of the problem, $\Phi_i(x,z)$ has to be an even function of $z$, and 
$\Phi_2(-\infty,z)-\Phi_1(\infty,z)=\Phi_D$ for any fixed position $z$ in the limit of large wall separations $L\rightarrow\infty$.
$\Phi_i^e(x,z)$ exhibits these properties.

By adding the electrostatic potentials of two single walls, each in contact with the fluid interface in a semi-infinite geometry
with respect to $z$,
one obtains the superposition approximation (denoted by superscript ``$s$'')
\begin{align}
   &\Phi_i^s(x,z) = \!2\Phi_{bi}\!+\!\!\sum\limits_{j\in\{1,2\}}^{j\neq i}
                      \frac{2(-1)^j\kappa_j\varepsilon_j\Phi_D}{\kappa_1\varepsilon_1+\kappa_2\varepsilon_2}
                      e^{-\kappa_{i}\lvert x\rvert}\notag\\
                     &+\ 2\Phi_i^{(0)} \cosh(\kappa_i z)e^{-\kappa_i L}\!\notag\\
                     &+\!\!\sum\limits_{j\in\{1,2\}}^{j\neq i}\int\displaylimits_0^{\infty} dq~C_{ij}^s(q) \cos(qL) \cos(qz)
                      e^{-\sqrt{q^2+\kappa_i^2}\lvert x\rvert}.
   \label{eq:m2}
\end{align}
The explicit expression for $C_{ij}^s(q)$ is given in appendix~\ref{app:A}. A comparison between the exact electrostatic 
potential $\Phi_i^e(x,z)$ and the superposition approximation $\Phi_i^s(x,z)$ at the plane of interface ($x=0$) is displayed 
in Fig.~\ref{fig:3si} in the appendix. Moreover, $\Phi_i^s(x,z)$ does not satisfy the boundary condition which relates 
the electric displacement field at the walls to the surface charge densities and
$\Phi_2^s(-\infty,z)-\Phi_1^s(\infty,z)\neq\Phi_D$ for any fixed position $z$ in the limit of large wall separations $L\rightarrow\infty$.

%===============================================================================

\section{Surface and line interactions} With the electrostatic potential given, the corresponding grand canonical potential 
can also be determined both exactly as well as within the superposition approximation.
After subtracting the bulk free energy, the surface and interfacial tensions, and the line tension contributions from the grand potential one 
obtains the $L$-dependent part of the grand potential,
\begin{align}
\Delta\Omega(L)= A_1\omega_{\gamma,1}(L) + A_2\omega_{\gamma,2}(L) + \ell\omega_\tau(L),
\label{eq:m3}
\end{align}
for the walls being a distance $2L$ apart, where $A_1$ and $A_2$ are the total areas of the two walls in contact with
medium ``1'' and ``2'', respectively, and $\ell$ is the total length of the three-phase contact lines formed by medium ``1'', medium 
``2'', and the walls; by construction $\Delta\Omega(L\rightarrow\infty)\rightarrow0$. 
The surface interaction energy per total surface area $A_i$ ($\omega_{\gamma,i}$) in contact with medium $i\in\{1,2\}$ is exactly (superscript ``e'') given by
\begin{align}
   \omega^e_{\gamma,i}(L) = \frac{\sigma_i^2}{2\kappa_i\varepsilon_i}\left(\coth(\kappa_iL)-1\right),
\label{eq:m4}
\end{align}
and within the superposition approximation (superscript ``s'') by
\begin{align}
   \omega^s_{\gamma,i}(L) = \frac{\sigma_i^2}{2\kappa_i\varepsilon_i}\left(2e^{-\kappa_iL}\cosh(\kappa_iL)-1\right).
\label{eq:m5}
\end{align}

According to Eqs.~\Eq{m4} and \Eq{m5}, varying $\sigma_i$ and $\varepsilon_i$ influences only the amplitude of $\omega_{\gamma,i}$ 
whereas its decay rate is solely determined by $\kappa_i$.
For large wall separations one has $\dps\omega_{\gamma,i}^e(\kappa_iL\gg1)\simeq\frac{\sigma_i^2}{\kappa_i\varepsilon_i}e^{-2\kappa_iL}$ 
and $\dps\omega_{\gamma,i}^s(\kappa_iL\gg1)\simeq\frac{\sigma_i^2}{2\kappa_i\varepsilon_i}e^{-2\kappa_iL}$, i.e., the superposition 
approximation correctly predicts the exponential decay in the large distance limit but, in contrast to common 
expectations, the corresponding prefactor is too small by a factor of $2$.
Moreover, the superposition approximation is qualitatively wrong for small wall separations (but still large on the molecular scale), 
because the exact surface interaction
potential diverges in this limit as $\dps\omega_{\gamma,i}^e(\kappa_iL\ll1)=\frac{\sigma_i^2}{2\kappa_i\varepsilon_i}
\left[\frac{1}{\kappa_iL}-1+\frac{\kappa_iL}{3}+\mathcal{O}((\kappa_iL)^3)\right]$, whereas the superposition approximation stays finite: 
$\dps\omega_{\gamma,i}^s(\kappa_iL\ll1)=\frac{\sigma_i^2}{2\kappa_i\varepsilon_i}\left[1-2\kappa_iL+\mathcal{O}((\kappa_iL)^2)\right]$. 
Thus the superposition approximation underestimates $\omega_{\gamma,i}$ for all $L$. Since for dilute aqueous electrolyte solutions of, 
e.g., $1\,\mathrm{mM}$ ($\approx0.0006\,\mathrm{nm^{-3}}$) ionic strength the Debye length ($1/\kappa_i\gtrsim10\,\mathrm{nm}$) is much larger 
than typical molecular size (e.g., $L=1\,\mathrm{nm}$), the exact surface interaction $\omega_{\gamma,i}^e(L)$
and the corresponding superposition approximation $\omega_{\gamma,i}^s(L)$ differ by at least one order of magnitude: 
$\omega_{\gamma,i}^e(L)/\omega_{\gamma,i}^s(L)\simeq1/(\kappa_iL)\gtrsim10$. 
Figure~2(a) displays a comparison between the exact result (black solid lines) and the superposition approximation (red dashed lines) for  
a set of typical experimental values for the ratios $\sigma=\sigma_2/\sigma_1$, $\kappa=\kappa_2/\kappa_1$, and 
$\varepsilon=\varepsilon_2/\varepsilon_1$.

The line interaction potential $\omega_\tau(L)$ per total length of the three-phase contact line between media ``1'' and ``2''
and the walls has been calculated from Eqs.~\Eq{m1} and \Eq{m2} (see appendix~\ref{app:C} for explicit expressions).
A comparison between the exact result $\omega_\tau^e(L)$ and the corresponding superposition approximation $\omega_\tau^s(L)$ is displayed
in Fig.~2(b). Similar to the surface interaction potentials, $\omega_\tau^s(L)$ differs significantly from the exact result $\omega_\tau^e(L)$
at small wall separations $2L$. For large values of $L$, its absolute value is too small by a factor of $2$, like the surface contribution.

%===============================================================================

\section{Discussion} By considering a slab geometry, we have investigated the electrostatic interaction between two 
colloidal particles at close proximity trapped at the interface of two immiscible electrolyte solutions. In our calculations, we have considered the 
charge density at the surface of the colloids to be constant, forming a boundary condition. However, in actual systems the situation is slightly different. 
When two particles approach each other the electrostatic potential becomes deeper in the region between the particles. Due to that certain charged molecular surface 
groups recombine in order to adjust the electrostatic potential. Such a process can better be described by a charge regulation model \cite{Rus89}. Keeping in mind the 
actual complexity of the system considered here, we briefly discuss the implications of charge regulation by focusing on a simpler system which consists of an
electrolyte between two charged walls without a liquid-liquid interface in between. For such a system, the electrostatic potential with a  
surface charge density $\sigma_{wi}(L)$ at the two walls (which is constant for any fixed $L$) is given by $\Phi_{wi}^e=\frac{\sigma_{wi}^e(L)}{\kappa_{wi}\varepsilon_{wi}}
\frac{\cosh{\kappa_{wi}z}}{\sinh{\kappa_{wi}L}}$ for the {\it{e}}xact calculation (see Eqs.~\Eq{7} and \Eq{8} in the appendix) and 
by $\Phi_{wi}^s=\frac{2\sigma_{wi}^s(L)}{\kappa_{wi}\varepsilon_{wi}}e^{-\kappa_{wi}L}\cosh{(\kappa_{wi}z)}$ 
within the {\it{s}}uperposition approximation (see the first terms in Eqs.~\Eq{36} and \Eq{37} in the appendix). Here the subscript ``$wi$'' stands for the 
system {\it{w}}ithout {\it{i}}nterface and the 
quantities $\sigma_{wi}$, $\kappa_{wi}$, and $\varepsilon_{wi}$ indicate, respectively, the surface charge density at the walls, the inverse Debye 
length, and the permittivity of the medium between the two planes in the absence of the horizontal interface. The dependence of the surface charge 
densities $\sigma_{wi}^e(L)$ and $\sigma_{wi}^s(L)$ on $L$ originates from the charge regulation (see appendix~\ref{app:E}). Inserting these expressions for 
the electrostatic potential into Eq.~\Eq{39p} in the appendix and using the fact that $D_x(\mathbf{r})$ vanishes in the absense of a liquid-liquid interface
as it is the case here, leads to the following surface interaction energies per total surface area of both walls:
\begin{align}
   \omega^e_{\gamma,wi}(L) = \frac{\left(\sigma_{wi}^e(L)\right)^2}{2\kappa_{wi}\varepsilon_{wi}}\left(\coth(\kappa_{wi}L)-1\right)
\label{eq:m6}
\end{align}
and
\begin{align}
   \omega^s_{\gamma,wi}(L) = \frac{\left(\sigma_{wi}^s(L)\right)^2}{2\kappa_{wi}\varepsilon_{wi}}\left(2e^{-\kappa_{wi}L}\cosh(\kappa_{wi}L)-1\right).
\label{eq:m7}
\end{align}
We note that Eqs.~\Eq{m6} and \Eq{m7} are identical to Eqs.~\Eq{m4} and \Eq{m5}, respectively, except the fact that here the surface charge density
varies with the thickness $L$ of the slab.

We discuss the two limiting cases of small and large $L$ separately. In the limit $\kappa_{wi}L\ll1$ one has $\sigma_{wi}^e(L)\simeq-\text{sign}(q)e\sqrt{2nKL}$ for 
the exact calculation (Eq.~\Eq{50} in the appendix) and $\sigma_{wi}^s(L)$ is constant for the superposition approximation (see
appendix~\ref{app:E}). $K$ (with units 1/volume) is the equilibrium constant for 
the association-dissociation reaction 
of the surface groups, $n$ denotes the total number of surface sites per cross-sectional area where a dissociation reaction can take
place, and $q$ is the valency of the solvated ions due to the dissociation reaction at the wall surface (appendix~\ref{app:E}). This implies
$\dps\omega_{\gamma,wi}^e(L\rightarrow0)=\frac{e^2nKL}{\kappa_{wi}\varepsilon_{wi}}
\left[\frac{1}{\kappa_{wi}L}-1+\frac{\kappa_{wi}L}{3}+\mathcal{O}((\kappa_{wi}L)^3)\right]$ which is nonzero for $L=0$. 
On the other hand, the nonzero and finite limiting value $\sigma_{wi}^s(L\rightarrow0)\neq0$ within the superposition approximation 
is clearly unphysical because the charge density is expected 
to decrease upon decreasing the inter-particle separation distance $L$. If by fiat, in order to avoid this unphysical feature, 
in Eq.~\Eq{m7} we replace $\sigma_{wi}^s(L)$ by $\sigma_{wi}^e(L)$, in the limit of small $L$ one finds 
$\dps\omega_{\gamma,wi}^s(L\rightarrow0)=\frac{e^2nKL}{\kappa_{wi}\varepsilon_{wi}}\left[1-2\kappa_{wi}L+\mathcal{O}((\kappa_{wi}L)^2)\right]$, which
vanishes for $L\rightarrow0$.
In the opposite limit, i.e., for $\kappa_{wi}L\gg1$, one finds 
$\dps\omega_{\gamma,wi}^e\simeq\frac{\left(\sigma_{wi}^e(L)\right)^2}{\kappa_{wi}\varepsilon_{wi}}e^{-2\kappa_{wi}L}$ 
and, by using the same replacement as above, 
$\dps\omega_{\gamma,wi}^s\simeq\frac{\left(\sigma_{wi}^e(L)\right)^2}{2\kappa_{wi}\varepsilon_{wi}}e^{-2\kappa_{wi}L}=\frac{\omega_{\gamma,wi}^e}{2}$ with 
$\sigma_{wi}^e(L)$ given by Eq.~\Eq{49} in the appendix. Thus for the
simple slab system without a liquid-liquid interface, but with charge regulation, the exact calculation and the superposition approximation are also in disagreement by a 
factor of 2 in the large separation limit and they differ qualitatively in the small separation limit. For the more complicated system with a liquid-liquid interface, 
we can expect these discrepancies to persist.

%===============================================================================

\section{Conclusion}

Within a continuum model of two parallel plates with two different electrolyte solutions in between forming a liquid-liquid interface, 
we have derived exact expressions for the electrostatic potential as well as for the effective surface and the line interaction potentials. The comparison 
between the exact results and the corresponding expressions within the superposition approximation reveals that the latter
underestimates these quantities qualitatively at short distances and quantitatively even at large distances.
Depending on the specific experimental system, the difference at small distances can be significant.
The issue whether the deviations at large distances persist for a spherical geometry is left for future investigations.
We expect our results to improve the description of the effective interaction between colloidal particles trapped at fluid interfaces,
which plays an important role, e.g., in the formation of two-dimensional colloidal aggregates.

%===============================================================================

\begin{acknowledgments}
Helpful discussions with Alois W\"{u}rger are gratefully acknowledged.
\end{acknowledgments}

%===============================================================================

\appendix

\section{\label{app:A}Electrostatic Potential} 

\subsection{Exact solution}
\begin{figure}[!t]
\centering\includegraphics[width=0.95\linewidth]{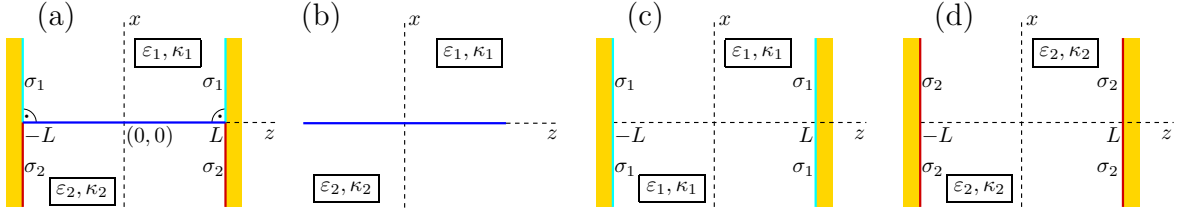}
\caption{(a) Slit of width $2L$ confined between two planar walls in the presence of an interface (solid blue line) between two 
electrolytes ``1'' and ``2'' characterized by permittivities $\varepsilon_1$, 
$\varepsilon_2$ and inverse Debye lengths $\kappa_1$, $\kappa_2$, respectively. In order to calculate the
electrostatic potential we divide the problem in three sub-problems: (i) two adjacent electrolytes ``1'' and ``2'' separated by an 
interface at $x=0$ in the absence of any walls (b), (ii) two homogeneous walls at $z=\pm L$ carrying charge densities $\sigma_1$ with the uniform 
electrolyte ``1'' in between (c), and (iii) two homogeneous walls at $z=\pm L$ carrying charge densities $\sigma_2$ with the uniform electrolyte
``2'' in between (d).}
\label{fig:1si}
\end{figure}

In order to obtain the electrostatic potential for the planar geometry shown in Fig.~\ref{fig:1}(b) (see also Fig.~\ref{fig:1si}(a)), the linearized 
Poisson-Boltzmann (Debye-H\"uckel) equation is solved in the two adjacent media with the following boundary conditions:
(i) the potential remains finite for $x\rightarrow\pm\infty$,
(ii) the electrostatic potential and the normal component of the electric displacement field are continuous at the fluid interface,
i.e., $\Phi_1(x=0^+,z)=\Phi_2(x=0^-,z)$ and $\varepsilon_1\partial_x\Phi_1(x=0^+,z)=\varepsilon_2\partial_x\Phi_2(x=0^-,z)$, and
(iii) the normal component of the electric displacement field at the walls correspond to the surface charge densities, i.e.,
$\varepsilon_i\partial_z\Phi_i(x,z=\pm L)=\pm\sigma_i$. In order to obtain such a solution we split the problem first into three 
sub-problems (Figs.~\ref{fig:1si}(b)-(d)): (i) only the fluid interface is present in the absence of any walls, 
(ii) two charged walls with homogeneous surface charge densities $\sigma_1$ and the uniform medium ``1'' in between, 
and (iii) two charged walls with homogeneous surface charge densities $\sigma_2$ 
and the uniform medium ``2'' in between.
The solution of sub-problem (i) will be denoted by $\bar\Phi_i(x)$ and the solutions of sub-problems (ii) and (iii) will be denoted
by $\Psi_1(z)$ and $\Psi_2(z)$, respectively.

\subsubsection{Solution of sub-problem (i)}

For two electrolyte solutions forming an interface at $x=0$ in the absence of any walls, the potential
can be calculated by solving
\begin{subequations}
\begin{align}
 \Delta\bar\Phi_1(x)-\kappa_{1}^2\bar\Phi_1(x)=0,~~~~~(x>0),
 \\
 \Delta(\bar\Phi_2(x)-\Phi_D)-\kappa_{2}^2(\bar\Phi_2(x)-\Phi_D)=0,~~~~~(x<0),
\end{align}
\label{eq:1}
\end{subequations}
\hspace{-0.3 cm} with $\Delta=d^2/dx^2$ and $\Phi_D$ denoting the Donnan potential (Galvani potential difference). The solutions of these equations can be written as
\begin{align*}
   \bar\Phi_1(x)=Ae^{-\kappa_{1}x}+Be^{\kappa_{1}x},&\\
          \bar\Phi_2(x)=\Phi_D+Ce^{-\kappa_{2}x}+De^{\kappa_{2}x}.&
\end{align*}
The boundary conditions $\bar\Phi_1(x\rightarrow\infty)\rightarrow0$ and $\bar\Phi_2(x\rightarrow-\infty)\rightarrow\Phi_D$ lead to $B=C=0$. The 
integration constants $A$ and $D$ can be obtained by using the boundary conditions that both the potential and the electric displacement field are 
continuous at the interface (i.e., at $x=0$). This leads to
\begin{subequations}
\begin{align}
 \bar\Phi_1(x)=\frac{\kappa_{2}\varepsilon_{2}\Phi_D}{\kappa_{1}\varepsilon_{1}+\kappa_{2}\varepsilon_{2}}e^{-\kappa_{1}x},
 \\
 \bar\Phi_2(x)=\Phi_D\left(1-\frac{\kappa_{1}\varepsilon_{1}}{\kappa_{1}\varepsilon_{1}+\kappa_{2}\varepsilon_{2}}e^{\kappa_{2}x}\right).
\end{align}
\end{subequations}

\subsubsection{Solutions of sub-problems (ii) and (iii)}

For two homogeneously charged walls at $z=\pm L$ with surface charge densities $\sigma_1$ and uniform medium ``1'' in between, the 
electrostatic potential is given by
\begin{equation*}
 \Delta\Psi_1(z)-\kappa_{1}^2\Psi_1(z)=0,
\label{eq:4}
\end{equation*}
where $\Delta=d^2/dz^2$. The solution of this equation reads
\begin{equation*}
 \Psi_1(z)=Ae^{-\kappa_{1}z}+Be^{\kappa_{1}z}.
\label{eq:5}
\end{equation*}
The integration constants $A$ and $B$ are determined by the boundary condition that the electric displacement field is equal 
to the charge density at the two walls. This leads to
\begin{align*}
   -\kappa_1Ae^{\kappa_{1}L}+\kappa_1Be^{-\kappa_{1}L}=-\frac{\sigma_1}{\varepsilon_1},&\\
          -\kappa_1Ae^{-\kappa_{1}L}+\kappa_1Be^{\kappa_{1}L}=\frac{\sigma_1}{\varepsilon_1},&
\label{eq:6}
\end{align*}
with the solution $A=B=\sigma_1/\left(2\kappa_1\varepsilon_1\sinh(\kappa_1L)\right)$ so that
\begin{equation}
 \Psi_1(z)=\frac{\sigma_1}{\kappa_1\varepsilon_1}\frac{\cosh(\kappa_1z)}{\sinh(\kappa_1L)}.
\label{eq:7}
\end{equation}
Sub-problem (iii) can be solved similarly leading to
\begin{equation}
 \Psi_2(z)=\frac{\sigma_2}{\kappa_2\varepsilon_2}\frac{\cosh(\kappa_2z)}{\sinh(\kappa_2L)}.
\label{eq:8}
\end{equation}

\subsubsection{Construction of a correction function and the final solution}

In view of the linear nature of the Debye-H\"uckel equation, one can add the solution of problem (ii) and the solution of problem (i) 
for the upper half-space, and the solution of problem (iii) and the solution of problem (i) for the lower half-space in order to obtain
solutions in each media which are also solutions of the Debye-H\"uckel equation. The sum $\bar\Phi_i(x)+\Psi_i(z)$ 
fulfills almost all boundary conditions for the electrostatic potential except continuity at the interface; although $\bar\Phi_i$
fulfills it, $\Psi_i$ violates it. In order to rectify this,
we construct a correction function $c_i(x,z)$ which has the following properties: (i) $c_i(x,z)$ is a solution of the Debye-H\"{u}ckel 
equation, i.e., $\Delta c_i(x,z)-\kappa_{i}^2 c_i(x,z)=0$, where $i=1~(2)$ corresponds to $x>0~(<0)$, (ii) $\partial_{z}c_i(x,z)=0$ 
at $z=\pm L$, (iii) $c_i(x=\pm\infty,z)=0$, (iv) $c_1(0^+,z)+\Psi_1(z)=c_2(0^-,z)+\Psi_2(z)$, and  
(v), due to $\partial_x\Psi_i(z)=0$, $\varepsilon_1\partial_{x}c_1(0^+,z)=\varepsilon_2\partial_{x}c_2(0^-,z)$.
It is clear from the construction that $c_i(x,z)$ keeps all the conditions, which are already satisfied by 
$\bar\Phi_i(x)+\Psi_i(z)$, unchanged and takes care of the continuity of the total potential at the interface. Therefore 
the exact electrostatic potential (superscript ``e'') is given by $\Phi_i^e(x,z)=\bar\Phi_i(x)+\Psi_i(z)+c_i(x,z)$.

In order to determine the correction function $c_i(x,z)$ we expand its dependence on $z\in[-L,L]$ into a Fourier series:
\begin{equation*}
 c_i(x,z)=\frac{a_{0,i}(x)}{2}+\sum\limits_{n=1}^\infty a_{n,i}(x)\cos\left(\frac{n\pi z}{L}\right)+\sum\limits_{n=1}^\infty b_{n,i}(x)\sin\left(\frac{n\pi z}{L}\right).
\label{eq:9}
\end{equation*}
The boundary condition $\partial_{z}c_i(x,\pm L)=0$ leads to $b_{n,i}(x)=0$, so that
\begin{equation}
 c_i(x,z)=\frac{a_{0,i}(x)}{2}+\sum\limits_{n=1}^\infty a_{n,i}(x)\cos\left(\frac{n\pi z}{L}\right).
\label{eq:10}
\end{equation}
Inserting this expression into the Debye-H\"{u}ckel equation (condition (i) listed above) one obtains
\begin{align*}
   \frac{a_{0,i}''(x)}{2}-\kappa_i^2\frac{a_{0,i}(x)}{2}+\sum\limits_{n=1}^\infty a_{n,i}''(x)\cos\left(\frac{n\pi z}{L}\right)& -\frac{n^2\pi^2}{L^2}\sum\limits_{n=1}^\infty a_{n,i}(x)\cos\left(\frac{n\pi z}{L}\right)\\
          & -\kappa_i^2\sum\limits_{n=1}^\infty a_{n,i}(x)\cos\left(\frac{n\pi z}{L}\right)=0,
\end{align*}
which implies
\begin{align*}
   a_{0,i}''(x)-\kappa_i^2a_{0,i}(x)=0,&\\
          a_{n,i}''(x)-\left[\left(\frac{n\pi}{L}\right)^2+\kappa_i^2\right]a_{n,i}(x)=0.&
\end{align*}
As solutions for these two equations one obtains
\begin{align*}
   a_{0,i}(x)=D_ie^{-\kappa_i x}+C_ie^{\kappa_i x},&\\
          a_{n,i}(x)=A_{n,i}e^{-\sqrt{\left(\frac{n\pi}{L}\right)^2+\kappa_i^2}x}+B_{n,i}e^{\sqrt{\left(\frac{n\pi}{L}\right)^2+\kappa_i^2}x}.&
\end{align*}
Due to the boundary condition $c_i(\pm\infty,z)=0$ the coefficients $a_{0,i}$ and $a_{n,i}$ in the two media are given by
\begin{align*}
   a_{0,1}(x)=D_1e^{-\kappa_1x},&\\
          a_{n,1}(x)=A_{n,1}e^{-\sqrt{\left(\frac{n\pi}{L}\right)^2+\kappa_1^2}x},&~~~~~\text{in medium 1,}
\end{align*}
and
\begin{align*}
   a_{0,2}(x)=C_2e^{\kappa_2x},&\\
          a_{n,2}(x)=B_{n,2}e^{\sqrt{\left(\frac{n\pi}{L}\right)^2+\kappa_2^2}x},&~~~~~\text{in medium 2.}
\end{align*}
With this Eq.~(\ref{eq:10}) can be written as $\left(D_1=D,~C_2=C,~A_{n,1}=A_n,~\text{and}~B_{n,2}=B_n\right)$
\begin{subequations}
\begin{align}
 c_1(x,z)=\frac{De^{-\kappa_1x}}{2}+\sum\limits_{n=1}^\infty A_ne^{-\sqrt{\left(\frac{n\pi}{L}\right)^2+\kappa_1^2}x}\cos\left(\frac{n\pi z}{L}\right),
 \\
 c_2(x,z)=\frac{Ce^{\kappa_2x}}{2}+\sum\limits_{n=1}^\infty B_ne^{\sqrt{\left(\frac{n\pi}{L}\right)^2+\kappa_2^2}x}\cos\left(\frac{n\pi z}{L}\right).
\end{align}
\end{subequations}
In order to determine the constants $A_n$, $B_n$, $C$, and $D$, the boundary conditions (iv) and (v) are used:
\begin{subequations}
\begin{align}
 \frac{2\sigma_1}{\kappa_1^2\varepsilon_1L}+D&=\frac{2\sigma_2}{\kappa_2^2\varepsilon_2L}+C,
\\
 \frac{2\sigma_1}{\kappa_1^2\varepsilon_1L}\frac{(-1)^n}{\left(1+\frac{n^2\pi^2}{\kappa_1^2L^2}\right)}+A_n&=\frac{2\sigma_2}{\kappa_2^2\varepsilon_2L}\frac{(-1)^n}{\left(1+\frac{n^2\pi^2}{\kappa_2^2L^2}\right)}+B_n,
\\
 -\varepsilon_1\kappa_1D&=\varepsilon_2\kappa_2C,
\\
 -\varepsilon_1\sqrt{\left(\frac{n\pi}{L}\right)^2+\kappa_1^2}A_n&=\varepsilon_2\sqrt{\left(\frac{n\pi}{L}\right)^2+\kappa_2^2}B_n.
\end{align}
\end{subequations}
Here we have used the relationships $\int\displaylimits_{-L}^L\cos\left(\frac{n\pi z}{L}\right)~dz=0$, 
$\int\displaylimits_{-L}^L\cos\left(\frac{n\pi z}{L}\right)\cos\left(\frac{m\pi z}{L}\right)~dz=L\delta_{n,m}$, and
$\int\displaylimits_{-L}^L\cosh\left(\kappa z\right)\cos\left(\frac{m\pi z}{L}\right)~dz=2\kappa(-1)^m\left[\kappa^2+\left(\frac{m\pi}{L}\right)^2\right]^{-1}\sinh(\kappa L)$
(Eq.~(2.671.4) in Ref.~\cite{Gra00}). Solving these four equations one finally arrives at the following expressions for 
the electrostatic potential in the two media:
\begin{align}  
					\Phi_1^e(x,z) &=\frac{\sigma_1}{\kappa_1\varepsilon_1}\frac{\cosh\kappa_1z}{\sinh(\kappa_1L)}+\frac{\kappa_{2}\varepsilon_{2}\Phi_D}{\kappa_{1}\varepsilon_{1}+\kappa_{2}\varepsilon_{2}}e^{-\kappa_{1}x}+\frac{1}{L}\frac{\frac{\sigma_2}{\kappa_2^2\varepsilon_2}-\frac{\sigma_1}{\kappa_1^2\varepsilon_1}}{1+\frac{\kappa_1\varepsilon_1}{\kappa_2\varepsilon_2}}e^{-\kappa_{1}x}\nonumber
 \\ 					&+2L\sum\limits_{n=1}^\infty(-1)^n\frac{\frac{\sigma_2}{\varepsilon_2}\frac{1}{n^2\pi^2+\kappa_2^2L^2}-\frac{\sigma_1}{\varepsilon_1}\frac{1}{n^2\pi^2+\kappa_1^2L^2}}{1+\frac{\varepsilon_1\sqrt{n^2\pi^2+\kappa_1^2L^2}}{\varepsilon_2\sqrt{n^2\pi^2+\kappa_2^2L^2}}}e^{-\sqrt{\left(\frac{n\pi}{L}\right)^2+\kappa_1^2}x}\cos\left(\frac{n\pi z}{L}\right), 
\label{eq:18}
\end{align}
and
\begin{align}  
					\Phi_2^e(x,z) &=\frac{\sigma_2}{\kappa_2\varepsilon_2}\frac{\cosh(\kappa_2z)}{\sinh(\kappa_2L)}+\Phi_D\left(1-\frac{\kappa_{1}\varepsilon_{1}}{\kappa_1\varepsilon_1+\kappa_2\varepsilon_2}e^{\kappa_2x}\right)+\frac{1}{L}\frac{\frac{\sigma_1}{\kappa_1^2\varepsilon_1}-\frac{\sigma_2}{\kappa_2^2\varepsilon_2}}{1+\frac{\kappa_2\varepsilon_2}{\kappa_1\varepsilon_1}}e^{\kappa_{2}x}\nonumber
 \\ 					&+2L\sum\limits_{n=1}^\infty(-1)^n\frac{\frac{\sigma_1}{\varepsilon_1}\frac{1}{n^2\pi^2+\kappa_1^2L^2}-\frac{\sigma_2}{\varepsilon_2}\frac{1}{n^2\pi^2+\kappa_2^2L^2}}{1+\frac{\varepsilon_2\sqrt{n^2\pi^2+\kappa_2^2L^2}}{\varepsilon_1\sqrt{n^2\pi^2+\kappa_1^2L^2}}}e^{\sqrt{\left(\frac{n\pi}{L}\right)^2+\kappa_2^2}x}\cos\left(\frac{n\pi z}{L}\right).
\label{eq:19}
\end{align}
Equations~(\ref{eq:18}) and (\ref{eq:19}) can be expressed in terms of a single equation:
\begin{align}
   \Phi_i^e(x,z)=\!\Phi_{bi}\!+\sum\limits_{j\in\{1,2\}}^{j\neq i}
                  \frac{(-1)^j\kappa_j\varepsilon_j\Phi_D}{\kappa_1\varepsilon_1+\kappa_2\varepsilon_2}
                  e^{-\kappa_{i}\lvert x\rvert}
                  &+\Phi_i^{(0)}\frac{\cosh(\kappa_iz)}{\sinh(\kappa_iL)}\!
                  +\!\!\sum\limits_{j\in\{1,2\}}^{j\neq i}\frac{C_{ij}^{(0)}(L)e^{-a_i^{(0)}(L)\lvert x\rvert}}{2}\notag\\
                 &+\sum\limits_{j\in\{1,2\}}^{j\neq i}\sum\limits_{n=1}^\infty C_{ij}^{(n)}(L)e^{-a_i^{(n)}(L)\lvert x\rvert}\cos\left(\frac{n\pi z}{L}\right),
   \label{eq:20}
\end{align}
with $\Phi_i^{(0)}=\sigma_i/(\varepsilon_i\kappa_i)$, $a_i^{(n)}(L)=\sqrt{\left(\frac{n\pi}{L}\right)^2+\kappa_i^2}$, $\Phi_{b1}=0$, $\Phi_{b2}=\Phi_D$, 
and
\begin{align}
C_{ij}^{(n)}(L)=2L(-1)^n\frac{\frac{\sigma_j}{\varepsilon_j}\frac{1}{n^2\pi^2+\kappa_j^2L^2}-\frac{\sigma_i}{\varepsilon_i}\frac{1}{n^2\pi^2+\kappa_i^2L^2}}{1+\frac{\varepsilon_i\sqrt{n^2\pi^2+\kappa_i^2L^2}}{\varepsilon_j\sqrt{n^2\pi^2+\kappa_j^2L^2}}}.
\label{eq:21}
\end{align}
Equation~(\ref{eq:20}) is identical to Eq.~(1) with the coefficients $C_{ij}^{(n)}(L)$ given by Eq.~(\ref{eq:21}).

We have checked that exactly the same result can be obtained by following the procedure adopted by Dom\'{\i}nguez \emph{et al.} \cite{Dom08}.

\subsection{Superposition approximation}

\begin{figure}[!t]
\centering\includegraphics[width=1.0\linewidth]{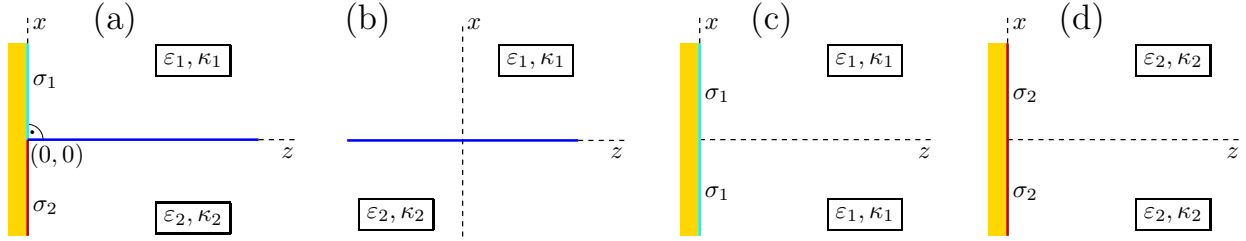}
\caption{ (a) Sketch of the problem with a {\it single} planar wall in the presence of a semi-infinite interface formed by electrolyte ``1'' and ``2'' 
with permittivities $\varepsilon_1$ and $\varepsilon_2$ as well as inverse Debye lengths $\kappa_1$ and $\kappa_2$, respectively. In order to 
facilitate the calculation of the
electrostatic potential the problem is sub-divided into three parts: (i) the electrolytes ``1'' and ``2'' separated by an interface at $x=0$ 
in the absence of any wall [(b)], (ii) a single homogeneous wall at $z=0$ carrying a charge density $\sigma_1$  bounding a uniform half-space filled by 
electrolyte ``1'' [(c)], and (iii) a single homogeneous wall at $z=0$ carrying a charge density $\sigma_2$ bounding a uniform half-space filled by 
electrolyte ``2'' [(d)].}
\label{fig:2si}
\end{figure}

First, we determine the electrostatic potential due to a {\it single} charged planar wall located at $z=0$ confining 
a semi-infinite interface 
between two electrolytes (Fig.~\ref{fig:2si}(a)). Also in this case we divide the problem into three sub-problems (Figs.~\ref{fig:2si}(b)-(d)): (i) a fluid interface only
in the absence of any wall, 
(ii) a homogeneously charged wall with surface charge density $\sigma_1$, bounding a half-space filled by uniform medium ``1'', and 
(iii) a homogeneously charged wall with surface charge 
density $\sigma_2$ bounding a half-space filled by uniform medium ``2''. After solving these three sub-problems a correction function is 
constructed which satisfies the following boundary conditions for the total electrostatic potential: 
(i) it is finite for $z\rightarrow\infty$ or $x\rightarrow\pm\infty$, 
(ii) the electrostatic potential and the normal component of the electric displacement field are continuous at the interface, and (iii) the 
normal component of the electric displacement field at the wall corresponds to the local surface charge density at the wall. 

\subsubsection{Solution of sub-problem (i)}

This part of the problem is identical to the sub-problem (i) we have considered for the exact solution. Thus the potentials in the two media are 
given by Eq.~(A2).

\subsubsection{Solutions of sub-problems (ii) and (iii)}

For a charged wall at $z=0$ carrying a charge density $\sigma_1$ in contact with the  uniform electrolyte ``1'', the electrostatic potential is given 
by the solution of
\begin{align*}
\Delta\Psi_1(z)-\kappa_{1}^2\Psi_1(z)=0,
\end{align*}
where $\Delta=d^2/dz^2$. The solution to this equation is given by
\begin{align*}
 \Psi_1(z)=Ee^{-\kappa_{1}z}+Fe^{\kappa_{1}z}.
\end{align*}
The boundary condition $\Psi_1(z\rightarrow\infty)\rightarrow0$ leads to $F=0$. In order to find the integration constant $E$ the boundary 
condition that the electric displacement field should be equal to the charge density at the wall, i.e., $-\varepsilon_1\partial_z\Psi_1(0)=\sigma_1$
is used. The final expression reads
\begin{align}
 \Psi_1(z)=\frac{\sigma_1}{\kappa_1\varepsilon_1}e^{-\kappa_{1}z}.
\label{eq:25}
\end{align}
Sub-problem (iii) can be solved analogously and the solution is given by
\begin{align}
 \Psi_2(z)=\frac{\sigma_2}{\kappa_2\varepsilon_2}e^{-\kappa_{2}z}.
\label{eq:26}
\end{align}

\subsubsection{Construction of the correction function and final solution}

We seek a correction function $c_i(x,z)$ such that (i) $c_i(x,z)$ is a solution of the Debye-H\"{u}ckel equation, i.e., 
$\Delta c_i(x,z)-\kappa_{i}^2 c_i(x,z)=0$ where $i=1~(2)$ corresponds to $x>0~(<0)$, (ii) $\partial_{z}c_i(x,z)=0$ at $z=0$, 
(iii) $c_i(x=\pm\infty,z)=c_i(x,z=\infty)=0$, 
(iv) $c_1(0^+,z)+\Psi_1(z)=c_2(0^-,z)+\Psi_2(z)$, and (v) $\varepsilon_1\partial_{x}c_1(0^+,z)=\varepsilon_2\partial_{x}c_2(0^-,z)$.
Accordingly the final solution for the electrostatic potential of a single wall in each medium is given by $\Phi_i(x,z)=\bar\Phi_i(x)+\Psi_i(z)+c_i(x,z)$. 
In order to determine this correction function we extend the 
system to $z\in[-\infty,\infty]$ and solve the Debye-H\"{u}ckel equation in the entire
space by taking the Fourier transform with respect to $z$ \cite{Sti61}. The second condition listed above is 
satisfied automatically because the system is symmetric about the plane $z=0$. If a solution for $z\in\mathbb{R}$ satisfies this boundary
condition at $z=0$, it is the solution looked for in the range $z>0$. Therefore we are looking for a solution of the equation
\begin{equation}
 \left(\partial_x^2+\partial_z^2-\kappa_i^2\right)c_i(x,z)=0,
\label{eq:27}
\end{equation}
with
\begin{equation}
 c_i(x,z)=\frac{1}{2\pi}\int\displaylimits_{-\infty}^{\infty} dq~\hat c_i(x,q)e^{iqz},~~~
 \hat c_i(x,q)=\int\displaylimits_{-\infty}^{\infty} dz~c_i(x,z)e^{-iqz},
\label{eq:28}
\end{equation}
or equivalently
\begin{equation}
 \left(\partial_x^2-q^2-\kappa_i^2\right)\hat c_i(x,q)=0.
\label{eq:29}
\end{equation}
For the two media ``1'' and ``2'', the solutions of Eq.~(\ref{eq:29}) which fulfill boundary condition (iii) can be written as
\begin{subequations}
\begin{align}
 \hat c_1(x,q)=&M_1(q)e^{-p_1x},~~~~~p_1>0,
 \\
 \hat c_2(x,q)=&M_2(q)e^{p_2x},~~~~~~p_2>0,
\end{align}
\label{eq:30}
\end{subequations}
with $p_i^2=q^2+\kappa_i^2$. In order to determine $M_1(q)$ and $M_2(q)$, boundary conditions (iv) and (v) are used. 
To apply the fourth condition the Fourier transforms $\hat\Psi_i(q)$ of $\Psi_i(z)=\frac{\sigma_i}{\kappa_i\varepsilon_i}e^{-\kappa_i\lvert z\rvert}$ 
(in line with the above symmetry argument, Eqs. (\ref{eq:25}) and (\ref{eq:26})) are needed:
\begin{align}
 \hat\Psi_1(q)=\frac{\sigma_1}{\kappa_1\varepsilon_1}\frac{2\kappa_1}{q^2+\kappa_1^2}=B_1\frac{2\kappa_1}{q^2+\kappa_1^2},\nonumber
 \\
 \hat\Psi_2(q)=\frac{\sigma_2}{\kappa_2\varepsilon_2}\frac{2\kappa_2}{q^2+\kappa_2^2}=B_2\frac{2\kappa_2}{q^2+\kappa_2^2},\nonumber
\end{align}
where $B_i=\sigma_i/\left(\kappa_i\varepsilon_i\right)$. Using these, boundary conditions (iv) and (v) lead to the following set of equations
\begin{align}
 \frac{2\kappa_1 B_1}{q^2+\kappa_1^2}+M_1(q)=&~\frac{2\kappa_2 B_2}{q^2+\kappa_2^2}+M_2(q),\nonumber
\\
 -\varepsilon_1 p_1 M_1(q)=&~\varepsilon_2 p_2 M_2(q).\nonumber
\end{align}
Solving this set of equations for $M_1(q)$ and $M_2(q)$ and inserting into Eqs.~(A17a) and (A17b) leads to the following expressions:
\begin{subequations}
\begin{align}
 c_1(x,z)=\frac{1}{\pi}\int_{-\infty}^{\infty} dq~\frac{\varepsilon_2p_2}{\varepsilon_1p_1+\varepsilon_2p_2}\left(\frac{-\kappa_1 B_1}{q^2+\kappa_1^2}+\frac{\kappa_2 B_2}{q^2+\kappa_2^2}\right)e^{-p_1x+iqz},
 \\
 c_2(x,z)=-\frac{1}{\pi}\int_{-\infty}^{\infty} dq~\frac{\varepsilon_1p_1}{\varepsilon_1p_1+\varepsilon_2p_2}\left(\frac{-\kappa_1 B_1}{q^2+\kappa_1^2}+\frac{\kappa_2 B_2}{q^2+\kappa_2^2}\right)e^{p_2x+iqz}.
\end{align}
\end{subequations}
Since apart from the factor $\exp(iqz)$ both integrands are even functions of $q$, one finds that indeed boundary condition (ii), i.e.,
$\partial_zc_i(x,z)=0$ for $z=0$ is fulfilled. Moreover, this symmetry allows one to write these expressions in terms of trigonometric 
functions so that one arrives at the following final expressions for the electrostatic potentials in the two media:
\begin{align}  
					\Phi_1(x,z) &=\frac{\sigma_1}{\kappa_1\varepsilon_1}e^{-\kappa_{1}z}+\frac{\kappa_{2}\varepsilon_{2}\Phi_D}{\kappa_{1}\varepsilon_{1}+\kappa_{2}\varepsilon_{2}}e^{-\kappa_{1}x}\nonumber
 \\ 
					&+\frac{2\varepsilon_2}{\pi}\int_0^{\infty} dq\frac{\sqrt{q^2+\kappa_2^2}\cos(qz)e^{-\sqrt{q^2+\kappa_1^2}x}}{\varepsilon_1\sqrt{q^2+\kappa_1^2}+\varepsilon_2\sqrt{q^2+\kappa_2^2}}\left(\frac{-\sigma_1}{\varepsilon_1(q^2+\kappa_1^2)}+\frac{\sigma_2}{\varepsilon_2(q^2+\kappa_2^2)}\right),
\label{eq:34}
\end{align}
and
\begin{align}  
					\Phi_2(x,z) &=\frac{\sigma_2}{\kappa_2\varepsilon_2}e^{-\kappa_{2}z}+\Phi_D\left(1-\frac{\kappa_{1}\varepsilon_{1}}{\kappa_{1}\varepsilon_{1}+\kappa_{2}\varepsilon_{2}}e^{\kappa_{2}x}\right)\nonumber
 \\ 
					&+\frac{2\varepsilon_1}{\pi}\int_0^{\infty} dq\frac{\sqrt{q^2+\kappa_1^2}\cos(qz)e^{\sqrt{q^2+\kappa_2^2}x}}{\varepsilon_1\sqrt{q^2+\kappa_1^2}+\varepsilon_2\sqrt{q^2+\kappa_2^2}}\left(\frac{\sigma_1}{\varepsilon_1(q^2+\kappa_1^2)}-\frac{\sigma_2}{\varepsilon_2(q^2+\kappa_2^2)}\right).
\label{eq:35}
\end{align}

Equations~(\ref{eq:34}) and (\ref{eq:35}) express the electrostatic potential in the two media due to a single charged plane located at $z=0$. 
The superposition approximation amounts to approximate the electrostatic potential between two charged walls at $z=\pm L$ by the sum of the
electrostatic potentials due to two identical charged walls at $z=-L$ and $z=+L$. This is
accomplished via shifting the potential $\Phi_i$ by $-L$ to left, by $+L$ to the right, reflecting the latter about its new position, and adding
the former and the latter (the superscript ``s'' indicates the solution obtained within the superposition approximation):
$\Phi_i^s(x,z)=\Phi_i(x,z+L)+\Phi_i(x,-(z-L))$ so that
\begin{align}  
					\Phi_1^s&(x,z) =\frac{2\sigma_1}{\kappa_1\varepsilon_1}e^{-\kappa_{1}L}\cosh(\kappa_1z)+\frac{2\kappa_{2}\varepsilon_{2}\Phi_D}{\kappa_{1}\varepsilon_{1}+\kappa_{2}\varepsilon_{2}}e^{-\kappa_{1}x}\nonumber
 \\ 
					&+\frac{4\varepsilon_2}{\pi}\int_0^{\infty} dq~\frac{\sqrt{q^2+\kappa_2^2}\cos(qL)\cos(qz)}{\varepsilon_1\sqrt{q^2+\kappa_1^2}+\varepsilon_2\sqrt{q^2+\kappa_2^2}}\left(\frac{-\sigma_1}{\varepsilon_1(q^2+\kappa_1^2)}+\frac{\sigma_2}{\varepsilon_2(q^2+\kappa_2^2)}\right)e^{-\sqrt{q^2+\kappa_1^2}x}
\label{eq:36}
\end{align}
and
\begin{align}  
					\Phi_2^s&(x,z) =\frac{2\sigma_2}{\kappa_2\varepsilon_2}e^{-\kappa_{2}L}\cosh(\kappa_2z)+2\Phi_D\left(1-\frac{\kappa_{1}\varepsilon_{1}}{\kappa_{1}\varepsilon_{1}+\kappa_{2}\varepsilon_{2}}e^{\kappa_{2}x}\right)\nonumber
 \\ 
					&+\frac{4\varepsilon_1}{\pi}\int_0^{\infty} dq~\frac{\sqrt{q^2+\kappa_1^2}\cos(qL)\cos(qz)}{\varepsilon_1\sqrt{q^2+\kappa_1^2}+\varepsilon_2\sqrt{q^2+\kappa_2^2}}\left(\frac{\sigma_1}{\varepsilon_1(q^2+\kappa_1^2)}-\frac{\sigma_2}{\varepsilon_2(q^2+\kappa_2^2)}\right)e^{\sqrt{q^2+\kappa_2^2}x}.
\label{eq:37}
\end{align}
Equations~(\ref{eq:36}) and (\ref{eq:37}) can be expressed by a single equation of the form
\begin{align}
   \Phi_i^s(x,z) = \!2\Phi_{bi}\!&+\!\!\sum\limits_{j\in\{1,2\}}^{j\neq i}
                      \frac{2(-1)^j\kappa_j\varepsilon_j\Phi_D}{\kappa_1\varepsilon_1+\kappa_2\varepsilon_2}
                      e^{-\kappa_{i}\lvert x\rvert}+\ 2\Phi_i^{(0)} \cosh(\kappa_i z)e^{-\kappa_i L}\!\notag\\
                     &+\!\!\sum\limits_{j\in\{1,2\}}^{j\neq i}\int\displaylimits_0^{\infty} dq~C_{ij}^s(q) \cos(qL) \cos(qz)
                      e^{-\sqrt{q^2+\kappa_i^2}\lvert x\rvert}
   \label{eq:38}
\end{align}
with $\Phi_i^{(0)}=\sigma_i/(\varepsilon_i\kappa_i)$, $\Phi_{b1}=0$, $\Phi_{b2}=\Phi_D$, and
\begin{align} 		
   C_{ij}^s(q)=\frac{4\varepsilon_j}{\pi}\frac{\sqrt{q^2+\kappa_j^2}}{\varepsilon_i\sqrt{q^2+\kappa_i^2}+\varepsilon_j\sqrt{q^2+\kappa_j^2}}
					\left(\frac{\sigma_j}{\varepsilon_j(q^2+\kappa_j^2)}-\frac{\sigma_i}{\varepsilon_i(q^2+\kappa_i^2)}\right).
\label{eq:39} 
\end{align}
Equation~(\ref{eq:38}) corresponds to (Eq.~(2)) with the coefficients $C_{ij}^s(q)$ given by Eq.~(\ref{eq:39}). 

\begin{figure}[!t]
\centering\includegraphics[width=0.95\linewidth]{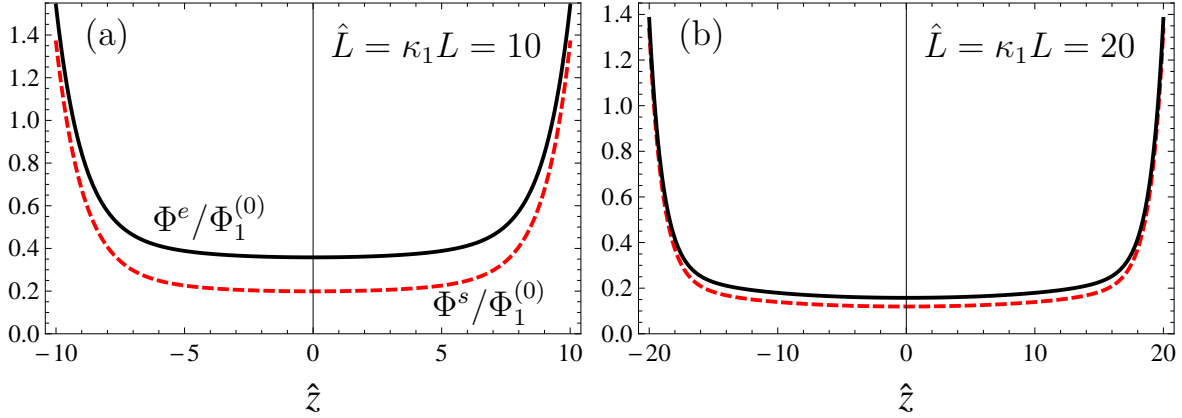}
\caption{Comparison between the exact expression (superscript ``e'', black solid lines, see Eq.~(1)) and the superposition approximation 
(superscript ``s'', red dashed lines, see Eq.~(2)) of the electrostatic potential $\Phi(x,z)$ at the interfacial plane ($x=0$) in units 
of $\Phi_1^{(0)}=\sigma_1/(\kappa_1\varepsilon_1)$ for varying $\hat{z}=\kappa_1z$ and two slit widths: 
$\hat{L}=\kappa_1L=10$ (panel (a)) and $\hat{L}=\kappa_1L=20$ (panel (b)).
For the plots typical parameter ratios $\kappa=\kappa_2/\kappa_1=0.025$, $\varepsilon=\varepsilon_2/\varepsilon_1=0.025$, 
$\sigma=\sigma_2/\sigma_1=0.1$, and $\Phi_D/\Phi_1^{(0)}=1.3$ have been chosen. $\Phi^e$ and $\Phi^s$ 
differ significantly at narrow widths $L$, and the difference between the two expressions decreases upon increasing the slit width. In the
limit $\hat{z}\rightarrow\pm\hat{L}$, both $\Phi^e$ and $\Phi^s$ remain finite.}
\label{fig:3si}
\end{figure}

A comparison between the exact and the approximate potential is given in Fig.~\ref{fig:3si}.

%-------------------------------------------------------------------------------

\section{\label{app:B}Grand potential}

\subsection{Density functional}
The model we are considering corresponds to the grand canonical density functional
\begin{align} 		
\beta\Omega\left[\rho_{\pm}\right]=\int\displaylimits_Vd^3r\left[\sum\limits_{i=\pm}\rho_i(\vec{r})\left\{\ln\left(\frac{\rho_i(\vec{r})}{\zeta_i}\right)-1+\beta V_i(\vec{r})\right\}+\frac{\beta \vec{D}\left(\vec{r},\left[\rho_\pm\right]\right)^2}{2\varepsilon(\vec{r})}\right],
\label{eq:39a} 
\end{align}
where `$+$' and `$-$' indicate the positive and negative ions respectively, $\beta=1/\left(k_BT\right)$ is the inverse thermal energy, $\rho_\pm(\vec{r})$ are 
the number densities of the ionic components, $\zeta_\pm$ represent the fugacities of the two ion-species, and $\varepsilon(\vec{r})$ denotes the permittivity 
with $\varepsilon(\vec{r})=\varepsilon_1~(\varepsilon_2)$ for $x>0~(x<0)$. Since the salt reservoir is provided by the bulk of media `1'
and `2', we use the freedom to shift the potentials $V_\pm(\vec{r})$, which describe the ion-solvent interactions due to solvation, such that
$V_\pm(\vec{r})=0$ in medium `1' ($x>0$) and $V_\pm(\vec{r})=f_\pm$ in medium `2' ($x<0$). Hence $f_{\pm}$ correspond to the ion solvation free energy
differences between media `2' and `1'. The integration volume $V$ is the slab formed in between the two charged 
planar walls. According to Gauss' law $\vec{\nabla}\cdot\vec{D}\left(\vec{r},\left[\rho_\pm\right]\right)=\sum\limits_{i=\pm}eq_i\rho_i(\vec{r})$ 
with $e>0$ the elementary charge and $q_{\pm}=\pm 1$. 
We consider Neumann-type boundary conditions at the walls, i.e., $\vec{n}(\vec{r})\cdot\vec{D}\left(\vec{r},\left[\rho_\pm\right]\right)=-\sigma(\vec{r})$ 
with the electric displacement field $\vec{D}$ and the charge density at the walls $\sigma(\vec{r})$. In Eq.~(\ref{eq:39a}) the sum
represents the entropic ideal gas contribution of the ions and the last term represents the energy contribution due to the electrostatic 
Coulomb interaction between the ions which is expressed in terms of the electrostatic energy density \cite{Iba13}. In this model the 
ions are pointlike particles.

\subsection{Expansion of the density functional}
Denoting the deviations of the ion number densities from the bulk ionic strength $I(\vec{r})$ ($I(\vec{r})=I_1$ for $x>0$ and $I(\vec{r})=I_2$ for $x<0$)
by $\phi_\pm(\vec{r}):=\rho_\pm(\vec{r})-I(\vec{r})$ and expanding the grand potential 
functional $\beta\Omega\left[\rho_{\pm}\right]$ in terms of the small deviations $\phi_i$ up to quadratic order one obtains
\begin{align*} 		
\beta\Omega\left[\rho_{\pm}\right]&=\int\displaylimits_Vd^3r\sum\limits_{i=\pm}I(\vec{r})\left[\ln\left(\frac{I(\vec{r})}{\zeta_i}\right)-1+\beta V_i(\vec{r})\right]\\
&+\int\displaylimits_Vd^3r\left[\sum\limits_{i=\pm}\phi_i(\vec{r})\left\{\ln\left(\frac{I(\vec{r})}{\zeta_i}\right)+\beta V_i(\vec{r})+\frac{\phi_i(\vec{r})}{2I(\vec{r})}\right\}+\frac{\beta \vec{D}\left(\vec{r},\left[\rho_\pm\right]\right)^2}{2\varepsilon(\vec{r})}\right]+\mathcal{O}(\phi^3). 
\end{align*}
Here the first line $(\mathcal{O}(\phi^0))$ describes the bulk contribution and the integrals in the second line 
$(\mathcal{O}(\phi^n);n\geq1)$ represent the surface and line contributions to the free energy (note that $\vec{D}=\mathcal{O}(\phi)$, 
see below). 
For future convenience we denote the latter by $\beta\mathcal{H}\left[\phi_\pm\right]$:
\begin{align} 		
\beta\mathcal{H}\left[\phi_\pm\right]=\int\displaylimits_Vd^3r\left[\sum\limits_{i=\pm}\phi_i(\vec{r})\left\{\ln\left(\frac{I(\vec{r})}{\zeta_i}\right)+\beta V_i(\vec{r})+\frac{\phi_i(\vec{r})}{2I(\vec{r})}\right\}+\frac{\beta \vec{D}\left(\vec{r},\left[\phi_\pm\right]\right)^2}{2\varepsilon(\vec{r})}\right].
\label{eq:39c} 
\end{align}

\subsection{Minimization of the density functional}
Minimization of $\beta\mathcal{H}\left[\phi_\pm\right]$ leads to the Euler-Lagrange equation $\delta\left(\beta\mathcal{H}\left[\phi_\pm\right]\right)=0$. 
Equation~(\ref{eq:39c}) implies
\begin{align} 		
\delta\left(\beta\mathcal{H}\left[\phi_\pm\right]\right)=\int\displaylimits_Vd^3r~\left[\sum\limits_{i=\pm}\delta\phi_i(\vec{r})\left\{\ln\left(\frac{I(\vec{r})}{\zeta_i}\right)+\beta V_i(\vec{r})+\frac{\phi_i(\vec{r})}{I(\vec{r})}\right\}\right]\nonumber\\
+\int\displaylimits_Vd^3r~\frac{\beta \vec{D}\left(\vec{r},\left[\phi_\pm\right]\right)}{\varepsilon(\vec{r})}\cdot\delta \vec{D}\left(\vec{r},\left[\phi_\pm\right]\right).
\label{eq:39d} 
\end{align}
Using the relation $\vec{D}\left(\vec{r},\left[\phi_\pm\right]\right)=-\varepsilon(\vec{r})\nabla\Phi\left(\vec{r},\left[\phi_\pm\right]\right)$, with 
$\Phi\left(\vec{r},\left[\phi_\pm\right]\right)$ denoting the electrostatic potential, and the divergence theorem, the last term in Eq.~(\ref{eq:39d}) 
can be written as
\begin{align*}
&\int\displaylimits_Vd^3r~\frac{\beta \vec{D}\left(\vec{r},\left[\phi_\pm\right]\right)}{\varepsilon(\vec{r})}\cdot\delta \vec{D}\left(\vec{r},\left[\phi_\pm\right]\right)\\
&=\int\displaylimits_Vd^3r~\beta\left(-\nabla\Phi\left(\vec{r},\left[\phi_\pm\right]\right)\right)\cdot\delta \vec{D}\left(\vec{r},\left[\phi_\pm\right]\right)\\
&=-\beta\int\displaylimits_{\partial V} d^2r~\Phi\left(\vec{r},\left[\phi_\pm\right]\right)\vec{n}(\vec{r})\cdot\delta\vec{D}\left(\vec{r},\left[\phi_\pm\right]\right)
+\beta\int\displaylimits_{V} d^3r~\Phi\left(\vec{r},\left[\phi_\pm\right]\right)\delta\nabla\cdot\vec{D}\left(\vec{r},\left[\phi_\pm\right]\right).
\end{align*}
The Neumann boundary condition leads to $\vec{n}(\vec{r})\cdot\delta\vec{D}\left(\vec{r},\left[\phi_\pm\right]\right)=-\delta\sigma(\vec{r})=0$. 
According to electrostatics one has $\nabla\cdot\vec{D}\left(\vec{r},\left[\phi_\pm\right]\right)=\sum\limits_{i=\pm}eq_i\rho_i(\vec{r})=\sum\limits_{i=\pm}eq_i\left(\phi_i(\vec{r})+I(\vec{r})\right)=\sum\limits_{i=\pm}eq_i\phi_i(\vec{r})$ (as $\sum\limits_{i=\pm}q_iI(\vec{r})=0$). 
This implies
\begin{align} 		
\delta\left(\beta\mathcal{H}\left[\phi_\pm\right]\right)=\int\displaylimits_Vd^3r~\sum\limits_{i=\pm}\delta\phi_i(\vec{r})\left[\ln\left(\frac{I(\vec{r})}{\zeta_i}\right)+\beta V_i(\vec{r})+\frac{\phi_i(\vec{r})}{I(\vec{r})}+\beta eq_i\Phi\left(\vec{r},\left[\phi_\pm\right]\right)\right].
\label{eq:39f} 
\end{align}
The Euler-Lagrange equation leads to
\begin{align} 		
\ln\left(\frac{I(\vec{r})}{\zeta_i}\right)+\beta V_i(\vec{r})+\frac{\phi_i(\vec{r})}{I(\vec{r})}+\beta eq_i\Phi\left(\vec{r},\left[\phi_\pm\right]\right)=0.
\label{eq:39g} 
\end{align}
We first discuss the bulk phases.

\subsubsection{Bulk of phase 1 ($x>0$)}

In the bulk of phase 1 one has $I(\vec{r})=I_1$, $\beta V_\pm(\vec{r})=0$, $\phi_\pm(\vec{r})=0$, and $\Phi\left(\vec{r},\left[\phi_\pm\right]\right)=0$.
Therefore Eq.~(\ref{eq:39g}) gives
\begin{align} 		
\ln\left(\frac{I_1}{\zeta_\pm}\right)=0
\label{eq:39h}
\end{align}
so that
\begin{align} 		
\zeta_\pm=I_1.
\label{eq:39h1}
\end{align}

\subsubsection{Bulk of phase 2 ($x<0$)}

In the bulk of phase 2 one has $I(\vec{r})=I_2$, $\beta V_\pm(\vec{r})=\beta f_\pm$, $\phi_\pm(\vec{r})=0$, 
and $\Phi\left(\vec{r},\left[\phi_\pm\right]\right)=\Phi_D$, where $\Phi_D$ is the 
Donnan potential (Galvani potential difference). Accordingly, Eq.~(\ref{eq:39g}) gives
\begin{align} 		
\ln\left(\frac{I_2}{\zeta_\pm}\right)+\beta f_\pm\pm\beta e \Phi_D=0.
\label{eq:39h2}
\end{align}
Using Eq.~(\ref{eq:39h1}) this can be written as
\begin{align} 		
\ln\left(\frac{I_2}{I_1}\right)+\beta f_\pm\pm\beta e \Phi_D=0.
\label{eq:39i}
\end{align}
Adding the two equations in Eq.~(\ref{eq:39i}), one obtains for the partition ratio
\begin{align} 		
2\ln\left(\frac{I_2}{I_1}\right)+\beta \left(f_++f_-\right)=0
\end{align}
so that
\begin{align} 		
\frac{I_2}{I_1}=\exp\left(-\frac{\beta}{2}\left(f_++f_-\right)\right).
\end{align}
Subtracting the two equations in Eq.~(\ref{eq:39i}) leads to the Donnan potential:
\begin{align} 		
\beta \left(f_+-f_-\right)+2\beta e\Phi_D=0
\end{align}
so that
\begin{align} 		
\Phi_D=-\frac{1}{2e}\left(f_+-f_-\right).
\end{align}
Combining Eqs.~(\ref{eq:39h}) and Eq.~(\ref{eq:39h2}) one can write:
\begin{align} 		
\ln\left(\frac{I(\vec{r})}{\zeta_\pm}\right)+\beta V_\pm(\vec{r})\pm\beta e \varphi(\vec{r})=0
\label{eq:39j}
\end{align}
with $\varphi(\vec{r})$ introduced such that $\varphi(\vec{r})=0$ for $x>0$ and $\varphi(\vec{r})=\Phi_D$ for $x<0$. Subtracting this bulk contribution 
from Eq.~(\ref{eq:39g}) one obtains
\begin{align} 		
\frac{\phi_i(\vec{r})}{I(\vec{r})}+\beta eq_i\left(\Phi\left(\vec{r},\left[\phi_\pm\right]\right)-\varphi(\vec{r})\right)=0
\end{align}
which can be rewritten as
\begin{align} 		
\phi_i(\vec{r})=-\beta eq_iI(\vec{r})\left(\Phi\left(\vec{r},\left[\phi_\pm\right]\right)-\varphi(\vec{r})\right).
\label{eq:39k}
\end{align}
With this Gauss' law gives
\begin{align} 		
-\nabla\cdot\left(\varepsilon(\vec{r})\nabla\Phi\left(\vec{r},\left[\phi_\pm\right]\right)\right)&=e\sum\limits_{i=\pm}q_i\phi_i(\vec{r})\nonumber\\
&=-2\beta e^2I(\vec{r})\left(\Phi\left(\vec{r},\left[\phi_\pm\right]\right)-\varphi(\vec{r})\right),
\label{eq:39l1}
\end{align}
The permittivity varies steplike as $\varepsilon(\vec{r})=\varepsilon_1\Theta(x)+\varepsilon_2\Theta(-x)$ where $\Theta$ is
the Heaviside step function. Using this Eq.~(\ref{eq:39l1}) can be written as
\begin{align} 		
\varepsilon_1\delta(x)\partial_x\Phi(\vec{r},\left[\phi_\pm\right])-\varepsilon_2\delta(x)\partial_x\Phi(\vec{r},\left[\phi_\pm\right])+\varepsilon(\vec{r})\nabla^2\Phi\left(\vec{r},\left[\phi_\pm\right]\right)=
2\beta e^2I(\vec{r})\left(\Phi\left(\vec{r},\left[\phi_\pm\right]\right)-\varphi(\vec{r})\right).
\label{eq:39l2}
\end{align}
For $x\neq0$ Eq.~(\ref{eq:39l2}) leads to
\begin{align} 		
\nabla^2\left(\Phi\left(\vec{r},\left[\phi_\pm\right]\right)-\varphi(\vec{r})\right)=\frac{2\beta e^2I(\vec{r})}{\varepsilon(\vec{r})}\left(\Phi\left(\vec{r},\left[\phi_\pm\right]\right)-\varphi(\vec{r})\right),
\label{eq:39m}
\end{align}
which is the linearized Poisson-Boltzmann equation
\begin{align}
 \nabla^2\left(\Phi\left(\vec{r},\left[\phi_\pm\right]\right)-\varphi(\vec{r})\right)=\kappa(\vec{r})^2
\left(\Phi\left(\vec{r},\left[\phi_\pm\right]\right)-\varphi(\vec{r})\right) 
\end{align}
with $\kappa(\vec{r})^2=2\beta e^2I(\vec{r})/\varepsilon(\vec{r})$. Integrating Eq.~(\ref{eq:39l2}) with respect to $x$ over the range $\left[-\alpha,\alpha\right]$ and taking $\alpha\rightarrow0$ leads to
the boundary condition of continuity of the electric displacement field at the interface: $\left(\varepsilon_1\partial_x\Phi(\vec{r})-\varepsilon_2\partial_x\Phi(\vec{r})\right)|_{x=0}=0$.

\subsection{Interaction potential}
The surface and line contributions to the free energy functional are given by Eq.~(\ref{eq:39c}).
Replacing therein $\left(\ln\left(I(\vec{r})/\zeta_i\right)+\beta V_i(\vec{r})\right)$ by $-\beta e q_i\varphi(\vec{r})$ according to Eq.~(\ref{eq:39j}), 
$\phi_i(\vec{r})/(2I(\vec{r}))$ by
$-\frac{\beta eq_i}{2}\left(\Phi\left(\vec{r},\left[\phi_\pm\right]\right)-\varphi(\vec{r})\right)$ according to Eq.~(\ref{eq:39k}) and using 
$\vec{D}(\vec{r},\left[\phi_\pm\right])=-\varepsilon(\vec{r})\nabla\Phi(\vec{r},\left[\phi_\pm\right])$ with 
$\nabla\cdot\vec{D}\left(\vec{r},\left[\phi_\pm\right]\right)=\sum\limits_{i=\pm}eq_i\phi_i(\vec{r})$ one can rewrite Eq.~(\ref{eq:39c}) as:
\begin{align} 		
\beta\mathcal{H}\left[\phi_\pm\right]=\int\displaylimits_Vd^3r~\left[-\frac{\beta}{2}\left(\nabla\cdot\vec{D}\left(\vec{r},\left[\phi_\pm\right]\right)\right)
\left(\Phi\left(\vec{r},\left[\phi_\pm\right]\right)+\varphi(\vec{r})\right)-\frac{\beta}{2}\vec{D}\left(\vec{r},\left[\phi_\pm\right]\right)\cdot\nabla\Phi\left(\vec{r},\left[\phi_\pm\right]\right)\right].
\end{align}
Using the product rule $\nabla\cdot(f\vec{F})=\nabla f\cdot\vec{F}+f\nabla\cdot \vec{F}$, where $f$ is a scalar and $\vec{F}$ is a vector, 
this can further be reduced to
\begin{align} 		
\mathcal{H}\left[\phi_\pm\right]=-\frac{1}{2}\int\displaylimits_Vd^3r~[\nabla\cdot\left\{\Phi\left(\vec{r},\left[\phi_\pm\right]\right)\vec{D}
\left(\vec{r},\left[\phi_\pm\right]\right)\right\}+\nabla\cdot\left\{\varphi(\vec{r})\vec{D}
\left(\vec{r},\left[\phi_\pm\right]\right)\right\}\nonumber\\
-\vec{D}\left(\vec{r},\left[\phi_\pm\right]\right)\cdot\nabla\varphi(\vec{r})].
\end{align}
Converting the volume integral into a surface integral by applying the divergence theorem and using the fact that $\nabla\varphi(\vec{r})=-\Phi_D\delta(x)\vec{e_x}$, 
one obtains
\begin{align} 		
\mathcal{H}\left[\phi_\pm\right]=-\frac{1}{2}\int\displaylimits_{\partial V}d^2r~\vec{n}(\vec{r})\cdot\vec{D}\left(\vec{r},\left[\phi_\pm\right]\right)\left(\Phi\left(\vec{r},\left[
\phi_\pm\right]\right)+\varphi(\vec{r})\right)-\frac{\Phi_D}{2}\int\displaylimits_{x=0}d^2r~D_x\left(\vec{r},\left[\phi_\pm\right]\right), 
\end{align}
where $D_x$ is the $x$ component of the electric displacement field $\vec{D}\left(\vec{r},\left[\phi_\pm\right]\right)$ and $x=0$ denotes the integration over the 
interfacial plane. Using the relation $\vec{n}(\vec{r})\cdot\vec{D}\left(\vec{r},\left[\phi_\pm\right]\right)=-\sigma(\vec{r})$ one finally arrives at the expression
\begin{align} 		
\mathcal{H}\left[\phi_\pm\right]=\frac{1}{2}\int\displaylimits_{\partial V}d^2r~\sigma(\vec{r})\left(\Phi\left(\vec{r},\left[
\phi_\pm\right]\right)+\varphi(\vec{r})\right)-\frac{\Phi_D}{2}\int\displaylimits_{x=0}d^2r~D_x\left(\vec{r},\left[\phi_\pm\right]\right).
\label{eq:39p} 
\end{align}
If the slab in between the charged planar walls is given by $V=[-L_x,L_x]\times[0,L_y]\times[-L,L]$ Eq.~(\ref{eq:39p}) can be written in
the following way (for brevity we skip the explicit functional dependence on $\phi_\pm$):
\begin{equation*}
\begin{split}
\mathcal{H}&=\frac{1}{2}\int\displaylimits_{\partial V}d^2r~\sigma(\vec{r})\left(\Phi(\vec{r})+\varphi(\vec{r})\right)-\frac{\Phi_D}{2}\int\displaylimits_{x=0}d^2r~D_x(\vec{r})\\
&=\frac{L_y}{2}\int\limits_{-L_x}^{L_x}dx~\left[\sigma(L)(\Phi(x,L)+\varphi(x))+\sigma(-L)(\Phi(x,-L)+\varphi(x))\right]-\frac{\Phi_D}{2}\int\limits_{x=0}d^2r~D_x(\vec{r})\\
&=\frac{L_y}{2}\int\displaylimits_{-L_x}^0dx\left[\sigma_2(L)(\Phi_2(x,L)+\Phi_D)+\sigma_2(-L)(\Phi_2(x,-L)+\Phi_D)\right]\\
&~~~+\frac{L_y}{2}\int\displaylimits_{0}^{L_x}dx\left[\sigma_1(L)\Phi_1(x,L)+\sigma_1(x,-L)\Phi_1(-L)\right]+\frac{\Phi_DL_y\varepsilon_1}{2}\int\displaylimits_{-L}^L dz~[\partial_x\Phi_1(x=0,z)],\\
\end{split}
\label{eq:39q}
\end{equation*}
where we have used $D_x(\vec{r})=-\varepsilon_1\partial_x\Phi_1(x=0,z)$, exploiting the continuity of the electric displacement field: 
$\varepsilon_1\partial_x\Phi_1(x=0,z)=\varepsilon_2\partial_x\Phi_2(x=0,z)$. For our system 
$\sigma_1(L)=\sigma_1(-L)=\sigma_1$, $\sigma_2(L)=\sigma_2(-L)=\sigma_2$, and the potentials in the two media are also symmetric with 
respect to the $z$-axis, i.e., $\Phi_1(L)=\Phi_1(-L)=\Phi_1$ and $\Phi_2(L)=\Phi_2(-L)=\Phi_2$. Accordingly, one can write
\begin{align}
\mathcal{H}=L_y\sigma_2\int\displaylimits_{-L_x}^0dx~\Phi_2(x,L)+L_y\sigma_1\int\displaylimits_{0}^{L_x}dx~\Phi_1(x,L)+\frac{\Phi_DL_y\varepsilon_1}{2}\int\displaylimits_{-L}^L dz~[\partial_x\Phi_1(x=0,z)]&\nonumber\\
+L_xL_y\sigma_2\Phi_D&.
\label{eq:39r}
\end{align}
Inserting the expressions for the electrostatic potentials $\Phi_1(x,z)$ and $\Phi_2(x,z)$ given by Eqs.~\Eq{m1} and \Eq{m2} one can determine the interaction 
potential from Eq.~(\ref{eq:39r}). It consists of five contributions: (i) the surface tensions acting between the 
charged walls and the adjacent fluids in contact (times their area of contact), (ii) the interfacial tension acting between the two fluids in contact at the plane 
$x=0$ (times the interfacial area), (iii) the line tension at the three-phase contact lines at both walls (times the total length of the 
three-phase contact lines), (iv) surface 
interaction energy densities ($\omega_{\gamma,i}(L)$) due to the effective interaction between the two charged walls (times the total surface area of the 
walls in contact with media $i\in{1,2}$), and (v) a line interaction energy density ($\omega_{\tau}(L)$) due to the effective interaction between the two three-phase 
contact lines (times the total length of the three-phase contact lines). The first three contributions are independent of the distance $2L$ between the 
two walls (note that although the interfacial tension is $L$-independent it is multiplied by the interfacial area which is proportional to $L$) 
whereas the last two contributions 
are $L$-dependent (expressed by $\Delta\Omega(L)$ in Eq.~\Eq{m3}). After identifying and separating all these terms, one arrives at the expressions for the surface 
interaction energy densities, given by Eqs.~(4) and (5), and for the line interaction energy densities, given by Eqs.~(\ref{eq:40}) and (\ref{eq:41}) below.

%-------------------------------------------------------------------------------

\section{\label{app:C}Line interaction potential}

The exact expression for the line interaction potential (see Eq.~(3)) is given by
\begin{align}
 \frac{\omega_\tau^e}{\omega_\tau^{(0)}}&=\frac{1}{2\hat{L}}\frac{\frac{2\sigma}{\kappa}-\kappa\varepsilon-\frac{\sigma^2}{\kappa^3\varepsilon}}{1+\kappa\varepsilon}\nonumber\\
&+\frac{1}{\hat{L}}\sum\limits_{n=1}^\infty\left[\frac{\frac{\sigma}{\varepsilon}\frac{1}{\frac{n^2\pi^2}{\hat{L}^2}+\kappa^2}-\frac{1}{\frac{n^2\pi^2}{\hat{L}^2}+1}}{1+\frac{\sqrt{\frac{n^2\pi^2}{\hat{L}^2}+1}}{\varepsilon\sqrt{\frac{n^2\pi^2}{\hat{L}^2}+\kappa^2}}}\frac{1}{\sqrt{\frac{n^2\pi^2}{\hat{L}^2}+1}}
+\frac{\frac{\sigma}{\frac{n^2\pi^2}{\hat{L}^2}+1}-\frac{\sigma^2}{\varepsilon}\frac{1}{\frac{n^2\pi^2}{\hat{L}^2}+\kappa^2}}{1+\frac{\varepsilon\sqrt{\frac{n^2\pi^2}{\hat{L}^2}+\kappa^2}}{\sqrt{\frac{n^2\pi^2}{\hat{L}^2}+1}}}\frac{1}{\sqrt{\frac{n^2\pi^2}{\hat{L}^2}+\kappa^2}}\right]\nonumber
\\
&-\int\limits_0^\infty dx~\left[\frac{\frac{\sigma}{\varepsilon}\frac{1}{x^2\pi^2+\kappa^2}-\frac{1}{x^2\pi^2+1}}{1+\frac{\sqrt{x^2\pi^2+1}}{\varepsilon\sqrt{x^2\pi^2+\kappa^2}}}\frac{1}{\sqrt{x^2\pi^2+1}}
+\frac{\frac{\sigma}{x^2\pi^2+1}-\frac{\sigma^2}{\varepsilon}\frac{1}{x^2\pi^2+\kappa^2}}{1+\frac{\varepsilon\sqrt{x^2\pi^2+\kappa^2}}{\sqrt{x^2\pi^2+1}}}\frac{1}{\sqrt{x^2\pi^2+\kappa^2}}\right]
\label{eq:40}
\end{align}
with $\omega_\tau^{(0)}=\sigma_1^2/(\kappa_1^2\varepsilon_1)$, $\hat{L}=\kappa_1 L$, $\sigma=\sigma_2/\sigma_1$, $\kappa=\kappa_2/\kappa_1$, and
$\varepsilon=\varepsilon_2/\varepsilon_1$. The difference between the infinite sum and the integral is such that in leading order 
for $\hat{L}\rightarrow\infty$ it cancels the first term $\sim1/\hat{L}$. Also the higher order terms in $(1/\hat{L})$ vanish so that $\omega_\tau^e$ decays 
exponentially for large $\hat{L}$. In the opposite limit, i.e., for $\hat{L}\rightarrow0$, $\omega_\tau^e$ diverges $\sim1/\hat{L}$.

Within the superposition approximation the line interaction potential is given by
\begin{align}
 \frac{\omega_\tau^s}{\omega_\tau^{(0)}}&=\frac{\kappa}{1+\varepsilon\kappa}\left(\frac{\sigma}{\kappa^2}-\varepsilon\right)\frac{\Phi_D}{2\Phi_1^{(0)}}\nonumber
\\
&-\frac{1}{\pi}\frac{\Phi_D}{\Phi_1^{(0)}}\int\limits_0^\infty d\hat{q}~\frac{\sqrt{\hat{q}^2+1}\sqrt{\hat{q}^2+\kappa^2}}{\sqrt{\hat{q}^2+1}+\varepsilon\sqrt{\hat{q}^2+\kappa^2}}\left[\frac{\sigma}{\hat{q}^2+\kappa^2}-\frac{\varepsilon}{\hat{q}^2+1}\right]\frac{\sin(2\hat{q}\hat{L})}{\hat{q}}\nonumber
\\
&+\frac{1}{\pi}\int\limits_0^\infty d\hat{q}~\frac{\sqrt{\hat{q}^2+\kappa^2}}{\sqrt{\hat{q}^2+1}+\varepsilon\sqrt{\hat{q}^2+\kappa^2}}\left[\frac{\sigma}{\hat{q}^2+\kappa^2}-\frac{\varepsilon}{\hat{q}^2+1}\right]\frac{\cos(2\hat{q}\hat{L})}{\sqrt{\hat{q}^2+1}}\nonumber
\\
&-\frac{1}{\pi}\int\limits_0^\infty d\hat{q}~\frac{\sqrt{\hat{q}^2+1}}{\sqrt{\hat{q}^2+1}+\varepsilon\sqrt{\hat{q}^2+\kappa^2}}\left[\frac{\sigma}{\hat{q}^2+\kappa^2}-\frac{\varepsilon}{\hat{q}^2+1}\right]\frac{\sigma}{\varepsilon}\frac{\cos(2\hat{q}\hat{L})}{\sqrt{\hat{q}^2+\kappa^2}},
\label{eq:41}
\end{align}
with the parameters $\Phi_D$, $\Phi_1^{(0)}$, $\hat{L}$, $\sigma$, $\varepsilon$, and $\kappa$ defined as above. It is important to note that, unlike 
${\omega_\tau^e}$, ${\omega_\tau^s}$ depends on the Donnan potential $\Phi_D$. This is due to the fact that the superposition potential $\Phi_i^s(x,z)$ 
does not satisfy the boundary condition which relates the electric displacement field at the walls to the surface charge densities. 
For $\hat{L}\rightarrow\infty$, the first, constant, term in Eq.~(\ref{eq:41}) is cancelled by the leading contribution of the
second term. In Eq.~(\ref{eq:41}) the third and the fourth term go to zero for $\hat{L}\rightarrow\infty$. 
Thus, as expected, in this limit $\omega_\tau^s$ vanishes. For $\hat{L}\rightarrow0$, the second term vanishes but all other terms remain nonzero. 
Accordingly, in this limit $\omega_\tau^s$ reaches a finite nonzero value.

%-------------------------------------------------------------------------------

\section{\label{app:D}Comparison between the expressions $\omega_{\gamma,1}^e$ and $\omega_{\gamma,1}^s$ for the effective surface interaction}

\begin{figure}[!h]
\centering\includegraphics[width=10cm]{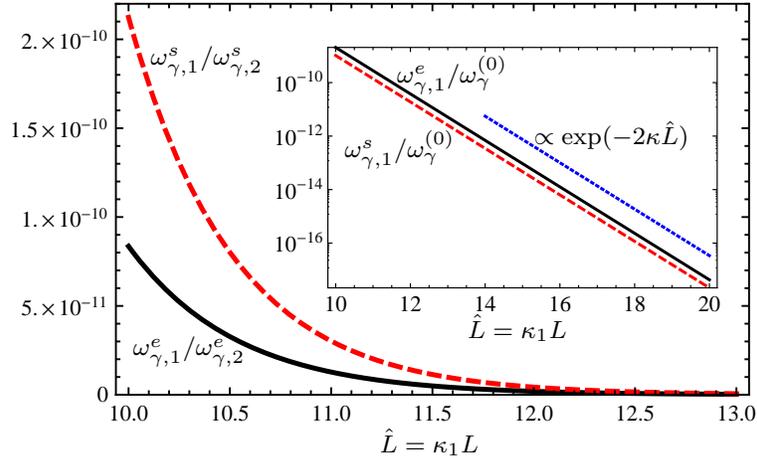}
\caption{ Comparison between the exact expression (superscript ``e'', black solid lines, see Eq.~\Eq{m4}) and the superposition 
approximation (superscript ``s'', red dashed lines, see Eq.~\Eq{m5}) for the surface interaction energy $\omega_{\gamma,1}(L)$ per 
total surface area of contact between the walls and medium ``1'' scaled by 
$\omega_{\gamma,2}(L)$ for varying $\hat{L}=\kappa_1L$.
Typical experimental values for the parameter ratios $\kappa=\kappa_2/\kappa_1=0.025$, $\varepsilon=\varepsilon_2/\varepsilon_1=0.025$, and 
$\sigma=\sigma_2/\sigma_1=0.1$ have been chosen for the plots. This data set is the same as the one used for Fig.~\ref{fig:2}, which 
displays the behavior of $\omega_{\gamma,2}^e$ and $\omega_{\gamma,2}^s$. Obviously $\omega^e_{\gamma,i}(L)$ and $\omega^s_{\gamma,i}(L)$ 
differ significantly at small separation distances, but even in the limit of large wall separations the superposition approximation is 
too small by a factor of $2$ (see the offset between the two curves in the inset).}
\label{fig:4si}
\end{figure}

%-------------------------------------------------------------------------------

\section{\label{app:E}Charge regulation model (In the absence of the interface)}

In the context of charge regulation we consider the reaction $AB\rightleftharpoons A^{-q}+B^q$ at the surface of the colloid, where $AB$ is the undissociated 
surface group which in the presence of the solvent dissociates into a charged surface site $A^{-q}$ and a solvated ion $B_q$ of valency $q$. We
consider the case that $B_q$ is one of the two ion species already present in the bulk electrolyte ($q=q_+=1$ if $B_q$ corresponds to the cation and
$q=q_-=-1$ if $B_q$ is the anionic species); the corresponding counterions of opposite charge are 
assumed not to contribute to the regulation of the surface charge. The equilibrium constant $K$ (with the unit 1/volume) for this reaction is given by
\begin{align}
 K=\frac{[A^{-q}]_s[B^q]_o}{[AB]_s},
 \label{eq:42}
\end{align}
where $[X]_s$ represents the number of species $X$ per surface area and $[Y]_o$ represents the number of species $Y$ per volume in the solution 
close to the surface \footnote{As we are considering only length scales larger than the bulk correlation length, $[Y]_o$ is obtained from the 
actual microscopic number density profile upon coarse graining, i.e., by averaging out its spatial variations at wavelengths up to the bulk correlation length.}. Then the surface charge density of the surface is
\begin{align}
 \sigma_{wi}=-qe[A^{-q}]_s,
 \label{eq:43}
\end{align}
the number of surface sites (dissociated plus undissociated) per cross-sectional area is
\begin{align}
 n=[A^{-q}]_s+[AB]_s,
 \label{eq:44}
\end{align}
and the number density of ions in the solvent close to the surface is given by
\begin{align}
 [B^q]_o=I+\phi_q(z=\pm L),
 \label{eq:45}
\end{align}
where $I$ is the bulk ionic strength (and as such independent of the dissociation reaction at the wall) and $\phi_q(z=\pm L)$ is the deviation close to the 
surface of the number density of ions of type $B$ from the bulk ionic strength. Since away from the walls the system considered here is homogeneous (due to the 
absence of the liquid-liquid interface), 
the quantities $I(\mathbf{r})$ and $\varphi(\mathbf{r})$ in Eq.~(\ref{eq:39k}) are constants. Setting $\varphi(\mathbf{r})=0$ without loss of generality, one has 
$\phi_q(z=\pm L)=-\beta qeI\Phi_p$, where $\Phi_p$ is the electrostatic potential at the {\it{p}}article surface, i.e., 
at $z=\pm L$. Using these, the dissociation constant can be written as
\begin{align}
 K=\frac{-\frac{\sigma_{wi}}{qe}I(1-\beta eq\Phi_p)}{n+\frac{\sigma_{wi}}{qe}}.
 \label{eq:46}
\end{align}
In the following we discuss the exact and superposition calculation separately.
\subsection{Exact calculation \label{ss1}}
In this case, the electrostatic potential at the walls is given by $\Phi_p=\frac{\sigma_{wi}^e}{\kappa_{wi}\varepsilon_{wi}\tanh(\kappa_{wi}L)}$ (see 
Eqs.~(\ref{eq:7}) and (\ref{eq:8})). According to Eq.~(\ref{eq:46}) this implies
\begin{align}
 K=\frac{-\sigma_{wi}^e I\left(1-\frac{\beta eq\sigma_{wi}^e}{\kappa_{wi}\varepsilon_{wi}\tanh(\kappa_{wi}L)}\right)}{\sigma_{wi}^e+qen}.
 \label{eq:47}
\end{align}
Solving this quadratic equation for $\sigma_{wi}^e$, one obtains,
\begin{align}
 \sigma_{wi}^e=\frac{\kappa_{wi}\varepsilon_{wi}\tanh(\kappa_{wi}L)}{2\beta eqI}\left(I+K\pm\sqrt{(I+K)^2+\frac{4\beta e^2q^2nIK}{\kappa_{wi}\varepsilon_{wi}\tanh(\kappa_{wi}L)}}\right).
 \label{eq:48}
\end{align}
Since $\sigma_{wi}^e\gtrless0$ for $q\lessgtr0$ and because the square root is larger than $(I+K)$, the negative sign in front of the square root has to be chosen. 
Using this and the expression for the inverse 
Debye length $\kappa_{wi}^2=2e^2I\beta/\varepsilon$, Eq.~(\ref{eq:48}) can be further simplified to
\begin{align}
 \sigma_{wi}^e=-\frac{e(I+K)\tanh(\kappa_{wi}L)}{q\kappa_{wi}}\left(\sqrt{1+\frac{2q^2n\kappa_{wi}K}{(I+K)^2\tanh(\kappa_{wi}L)}}-1\right).
 \label{eq:49}
\end{align}
For $L\rightarrow0$ this leads to
\begin{align}
 \sigma_{wi}^e(L\rightarrow0)\simeq-\text{sign}(q)e\sqrt{2nKL}.
 \label{eq:50}
\end{align}
This is the expression we have used in our discussions. As expected, $\lvert\sigma_{wi}^e(\kappa_{wi}L)\rvert$ decreases upon decreasing $L$.
\subsection{Superposition calculation}
Within the superposition approximation, the electrostatic potential at the {\it{p}}article surface is given by $\Phi_p=\frac{\sigma_{wi}^s}{\kappa_{wi}\varepsilon_{wi}}
\left(1+e^{-2\kappa_{wi}L}\right)$ (see the first terms in Eqs.~(\ref{eq:36}) and (\ref{eq:37})) so that according to Eq.~(\ref{eq:46})
\begin{align}
 K=\frac{-\sigma_{wi}^s I\left(1-\frac{\beta eq\sigma_{wi}^s\left(1+e^{-2\kappa_{wi}L}\right)}{\kappa_{wi}\varepsilon_{wi}}\right)}{\sigma_{wi}^s+qen}.
 \label{eq:51}
\end{align}
Solving this we obtain
\begin{align}
 \sigma_{wi}^s=\frac{\kappa_{wi}\varepsilon_{wi}}{2\beta eqI\left(1+e^{-2\kappa_{wi}L}\right)}
 \left(I+K\pm\sqrt{(I+K)^2+\frac{4\beta e^2q^2nIK\left(1+e^{-2\kappa_{wi}L}\right)}{\kappa_{wi}\varepsilon_{wi}}}\right).
 \label{eq:52}
\end{align}
As in Eq.~(\ref{eq:48}), in Eq.~(\ref{eq:52}) the negative sign in front of the square root has to be chosen. For $L\rightarrow0$ this attains a nonzero constant 
which is at odds with the expected behavior (see Sec. \ref{ss1} above). Thus for our discussion, 
instead of using Eq.~(\ref{eq:52}), we resort to Eqs.~(\ref{eq:49}) and (\ref{eq:50}) which offer a physically reasonable description of the dependence
of the charge density on $L$.

%===============================================================================

%===============================================================================

\end{document}